\documentclass[aps,prd,preprint,nofootinbib,superscriptaddress,a4paper]{revtex4-1}

\usepackage{graphicx,graphics,epsfig}

\usepackage{dcolumn}
\usepackage{feynmf}
\usepackage{verbatim}
\usepackage{subfigure}
\usepackage{amsmath,amssymb,amsbsy,multirow,wasysym,bm,bbm}
\usepackage{slashed}

\usepackage{array,booktabs,colortbl,colordvi,multirow}
\usepackage{colordvi,color,xcolor}
\usepackage{rotating}

\newcommand{\be}{\begin{equation}}
\newcommand{\ee}{\end{equation}}
\newcommand{\bea}{\begin{eqnarray}}
\newcommand{\eea}{\end{eqnarray}}
\newcommand{\bwt}{\begin{widetext}}
\newcommand{\ewt}{\end{widetext}}

\begin{document}
\title{Long-lived Colored Scalars at the LHC}

\author{Alejandro de la Puente}
\email[]{apuente@physics.carleton.ca}
\affiliation{Department of Physics, Carleton University, 1125 Colonel By Drive, Ottawa, ON K1S 5B6 Canada }
\affiliation{Theory Department, TRIUMF, 4004 Wesbrook Mall, Vancouver, B.C., Canada }

\author{Alejandro Szynkman}
\email[]{szynkman@fisica.unlp.edu.ar}
\affiliation{IFLP, CONICET -- Dpto. de F\'{\i}sica, Universidad Nacional de La Plata, C.C. 67, 1900 La Plata}

\date{\today}

\begin{abstract}
We study the collider signatures of a long-lived massive colored scalar transforming trivially under the weak interaction and decaying within the inner sections of a detector such as ATLAS or CMS. In our study, we assume that the colored scalar couples at tree-level to a top quark and a stable fermion, possibly arising from a dark sector or from supersymmetric extensions of the Standard Model. After implementing the latest experimental searches for long-lived colored scalars, we observe a region of parameter space consistent with a  colored electroweak-singlet scalar with mass between $\sim200-350$ GeV and a lifetime between $0.1-1$ $\text{mm}/c$ together, with a nearly degenerate dark fermion that may be probed at the $\sqrt{s}=13$ TeV LHC. We show that a search strategy using a combination of cuts on missing transverse energy and impact parameters can exclude regions of parameter space not accessed by prompt searches. We show that a region of parameter space within our simplified model may naturally arise from the light-stop window regime of supersymmetric extensions of the Standard Model, where a light mostly right-handed stop has a mass slightly larger than the lightest neutralino and decays through a four-body process. 

\end{abstract}

\maketitle

\section{Motivation}

 Searching for long-lived massive particles is an active program at the LHC. The ATLAS and CMS collaborations have carried out an extensive list of analysis targeting different topologies for the decay products of long-lived particles within the inner detector as well as particles with large enough lifetimes to escape the detector. In particular, both collaborations have carried out an analysis sensitive to very long-lived charged particles using time-of-flight (TOF) and energy loss, $dE/dx$ information~\cite{Chatrchyan:2013oca,ATLAS:2014fka}. This class of searches is highly model independent since a main requirement for an event to pass the experimental cuts is the appearance of a muon-like object in the muon spectrometer matched to a track in the inner detector. These searches rule out very long-lived gluinos, stops, sbottoms, and staus with masses below $\sim1250,900,850,500$ GeV respectively. Searches that target signatures in the inner detector have also been dealt with at the LHC. In particular, both CMS and ATLAS have conducted an analysis looking for disappearing tracks~\cite{CMS:2014gxa,Aad:2013yna}. This search is particularly interesting since it targets supersymmetric (SUSY) models where a chargino is near mass-degenerate with the lightest neutralino, rendering the chargino with a long lifetime. Within this scenario, the chargino decays to a neutralino and a charged pion with very low $p_{T}$, low enough for its track to pass reconstruction algorithms. The limits on the chargino mass are on masses below $\sim270$ GeV and proper lifetimes of $\sim0.2$ ns. Other searches look for interesting topologies that arise within SUSY scenarios with $R$-parity violation. In~\cite{CMS:2014wda,TheATLAScollaboration:2013yia}, the CMS and ATLAS collaborations search for long-lived neutral particles that decay to quark-antiquark pairs and a muon. These long-lived exotic particles are pair produced via the decay of a non-SM Higgs boson or the decay of a scalar quark to the lightest neutralino and a light quark. In the latter case the neutralino is long-lived and decays to a muon and a quark-antiquark pair through a small $R$-parity violating coupling. This search is sensitive to long-lived particles with a proper decay length between $2-40$ cm and production cross sections below $0.5-3$fb. A similar search is carried out by CMS~\cite{CMS:2014hka}, and targets scalar quark decays to a long-lived neutralino that subsequently decays to a same-flavor lepton pair and a neutrino with proper decay lengths in the range $0.01-100$ mm. The analysis is able to set an upper limit on the production cross section in the range $0.2-5$ fb for scalar quarks masses above $350$ GeV. In addition, the CMS collaboration has a very rigid search for displaced supersymmetry with a dilepton final state~\cite{Khachatryan:2014mea}, in this case one muon and one electron with opposite charges. The search is sensitive to the decay of a long-lived scalar top quark to a lepton and a b-jet in models of $R$-parity violation. The search has a highest mass exclusion of scalar top quark with mass below $790$ GeV and a proper lifetime of $2~\text{cm}/c$. The ATLAS collaboration has performed a very comprehensive study on displaced vertices, targeting long-lived particles decaying into two leptons or five or more charged particles~\cite{ATLASnew}. Limits are set on models of $R$-parity violation, split SUSY and gauge mediated SUSY breaking. In addition, searches have been conducted to analyze scenarios where a long-lived particle is stopped in the calorimeters and decays out of time~\cite{stGluinos,Aad:2013gva}. This class of analyses are sensitive to colored particles that hadronize and interact with matter, and strongly depend on the hadronization models implemented as well as the modeling of the hadron's interactions with matter. Displaced vertices are well motivated scenarios of models beyond the Standard Model. They appear within the context of SUSY breaking~\cite{Chen:1998awa,Randall:1998uk,Thomas:1998wy}, models of weak-scale $R$-parity violation~\cite{Hewett:2004nw,Barbier:2004ez,Graham:2012th}, Hidden valleys~\cite{Han:2007ae,Falkowski:2010cm,Falkowski:2010gv,Chan:2011aa}, baryogengesis~\cite{Barry:2013nva,Cui:2014twa}, dark QCD~\cite{Schwaller:2015gea} and late decays of right-handed neutrinos within SUSY extensions of the Standard Model~\cite{Cerdeno:2013oya}. In addition, there is an increasing activity in the recasting front, to better constrain a wide range of these models using the existing LHC displaced vertex analyses~\cite{Cui:2014twa,Schwaller:2015gea,Liu:2015bma}.
 
The analyses described above, which target long-lived massive particles in the inner detector, do not implement hard cuts on the missing transverse energy, $\slashed{E}_{T}$, to identify the signal and discriminate the background. This is due to the fact that either, one has a very compressed spectrum where there is very little missing energy carried away by a stable particle, or the decay products of the long-lived massive particles can all be reconstructed within the tracker. In addition, they are mostly driven by very specific models. In this work we would like to address whether a long-lived charged massive particle can leave a signature that can be reconstructed within the inner detector in association with a large amount of missing transverse energy, $\slashed{E}_{T}$. Furthermore, we wish to address this question in a model independent manner that can then be used as a template to compare experimental analyses to different theoretical frameworks. Due to the large number of experimental signatures and classes of long-lived massive particles, we focus only in the case where a colored electroweak-singlet scalar decays to a right-handed top quark and a dark fermion that is stable throughout the detector length. The dark fermion is a true source of missing energy, and depending on the parameters, it can carry away a large amount of energy. In our analysis we focus on pair production of the colored electroweak-singlet scalar and the semi-leptonic decay mode of the top quark. Our experimental signature would consist of a lepton pair with opposite charges, jets and missing energy; and in this way, our experimental signature is similar to the analysis on displaced SUSY by the CMS collaboration~\cite{Khachatryan:2014mea}, but we allow for the possibility of a large amount of $\slashed{E}_{T}$. 

Light colored scalar particles with the quantum numbers of an up-type quark are well motivated scenarios. This is the case within SUSY extensions of the Standard Model such as the Minimal Supersymmetric Standard Model (MSSM). Within this theoretical framework scalar top quarks are used to address the electroweak hierarchy problem of the SM. In addition, a light scalar top quark can be used to explain the observed baryon asymmetry, in particular, if its mass lies below that of the top quark. Recently, light stops have been studied within the framework of various SUSY breaking mechanisms and flavor structures~\cite{Brummer:2013upa,Kang:2012ra}, and various analysis have been dedicated to the feasibility of electroweak baryogenesis (EWB)~\cite{Delgado:2014kqa,Carena:2012np,Curtin:2012aa,Delgado:2012rk}. In addition, various phenomenological studies have looked at the possible collider signals of a light stop at the LHC. The work by the authors in~\cite{Morrissay} addresses all possible stop decay modes with special emphasis on the four-body and flavor violating channels, while the work by the authors in~\cite{Delgado} address primarily the four body decay mode. Recently the authors in~\cite{Ferretti:2015dea} have proposed a monojet-like search optimized for the four-body decay mode of the stop with an additional $b$-jet requirement. These studies significantly complement the search efforts by the CMS~\cite{Chatrchyan:2013xna,CMS:2014yma} and ATLAS~\cite{Aad:2014kra,Aad:2014nra} collaborations to address the light-stop window regime with the full data set at center of mass energies of $8$ TeV. In our study we aim to further complement the analysis of the light stop window, that is, with slightly displaced vertices in the range $0.1-10$ mm. 

The summary of our study is as follows: In Section~\ref{sec:Model} we introduce the model and in Section~\ref{sec:LLS} we study experimental constraints from direct searches at colliders, both prompt and long-lived. In Section~\ref{sec:LHC13} we present a search strategy that can be used to set limits on our framework at the LHC with $13$ TeV center of mass energies. In Section~\ref{sec:realizations} we discuss a model that may lead to our experimental signature and in Section~\ref{sec:discussion} we provide concluding remarks.

\section{Toy Model}\label{sec:Model}

In this section we introduce a toy model aimed at parametrizing the decay of a long-lived colored electroweak-singlet scalar, $\phi$, to a top quark and a new hidden Majorana fermion, $X$. The Majorana fermion communicates with the SM via the colored scalar. This new Majorana fermion is stable at scales larger than those of a modern detector and will be a source of true missing energy. We can write the coupling of the hidden sector to the SM top quark via the following Lagrangian:
\begin{equation}
-{\cal L}=\lambda\phi^{*}\bar{X}P_{R}t+~\text{h.c}.\label{eq:toy}
\end{equation}

Models featuring a Majorana fermion, singlet under the SM gauge group and a non-trivially charged scalar have been studied in~\cite{Arcadi:2014tsa} in the context of minimal decaying dark matter. In particular, they have explored the possibility that the stability of the dark matter candidate, $X$, is not guaranteed. The study covers a wide range of parameter space in accordance with a freeze-in mechanism~\cite{Hall:2009bx}, arising from the decays of the colored scalars in thermal equilibrium, and a superWIMP mechanism~\cite{Feng:2003xh}, from the decay of the colored scalar after it has undergone freeze out, to account for the observed dark matter relic abundance. In our study we limit our analysis to a stable Majorana fermion, stable throughout the length of a modern particle collider detector such as ATLAS or CMS and look further into its identity in Section~\ref{sec:realizations}.

In the limit where the mass of the colored electroweak-singlet scalar, $m_{\phi}$, is larger than $m_{t}+m_{X}$, the leading decay mode is given by
\begin{eqnarray}
\Gamma_{\phi\to t X}&=& \frac{1}{16\pi m^{3}_{\phi}} \lambda^{2}\left[ m_{\phi}^2-m_{t}^2-m_{X}^2\right] \left[m_{\phi}^{4}+m^{4}_{t}+m^{4}_{X}-2(m^{2}_{\phi}m^{2}_{t}+m^{2}_{\phi}m^{2}_{X}+m^{2}_{t}m^{2}_{X})\right]^{1/2},\nonumber \\ 
\end{eqnarray}
whereas for masses below $m_{t}+m_{X}$, the three-body decay mode through an off-shell top and the four-body mode through and off shell $W$ open up. An analytic expression for the three-body decay mode in the context of the MSSM where a stop decays to a $W$ gauge boson, a $b$-quark and a neutralino has been studied in~\cite{Porod:1998kk}. The four-body decay width can be calculated using the approximate formula introduced in~\cite{Delgado} and its validity is for small $\Delta M=m_{\phi}-m_{X}$. In particular, for masses $m_{\phi}-m_{X}<m_{W}$ the four-body decay mode branching fractions are given by
\begin{eqnarray}
Br(\phi\to bjj' X)&\approx& 1/3 \nonumber \\
Br(\phi\to bl\nu_{l} X) &\approx& 1/9.
\end{eqnarray}
Furthermore, in this class of theories, a flavor violating decay mode may exist and compete with the four body decay channel. SUSY theories are one classic example where the colored electroweak-singlet scalar identified with a scalar top quark decays into a charm quark and the lightest neutralino. This coupling can be parametrized by a mixing angle, $\theta_{t,c}$ which depends on the flavor structure of the model. However, the authors in ~\cite{Delgado,Morrissay} show that this region is very complex as the four body and flavor violating channels dominate in different regions but become comparable for a mixing parameter of order $10^{-5}$ or when the stop is mostly right-handed. The study in~\cite{Morrissay} carries out an in-depth analysis of the sensitivity that the LHC has to the different decay modes of a light stop within the minimal SUSY Standard Model and within the context of a promptly decaying stop. The authors in~\cite{Delgado} instead, focus primarily on a prompt mostly right-handed stop decaying to leptons through a four-body decay channel. In both cases it appears that a light stop is only viable if its mass lies above $200$ GeV. 

Within our framework, two colored electroweak-singlets can be pair produced, each decaying to a $b$-quark, a lepton and missing energy. For small couplings, $\lambda$, and mass difference $\Delta M=m_{\phi}-m_{X}$, $\phi$ may be long-lived and decay within the inner detector, particularly for proper decay lengths in the range $0.1-100$ mm. The leptons can be reconstructed and the long lifetime of $\phi$ may be inferred by the displacement of the leptons from the primary vertex through their impact parameters. This is due to the large correlation that exists between the unstable particle's lifetime and the impact parameters of the decay products. The analysis by the CMS collaboration~\cite{Khachatryan:2014mea} is especially sensitive to such a signature with the exception that the analysis is performed without the use of hard cuts on the hadronic activity or missing transverse energy. In what follows we study the sensitivity that the LHC has to our simplified model, in particular, to scenarios with large $\slashed{E}_{T}$.

\section{Long-lived Colored Scalars at LHC 8 TeV}\label{sec:LLS}
\subsection{Prompt searches at LHC 8 TeV}

\begin{figure*}[ht]
\centering
\includegraphics[width=10.0cm]{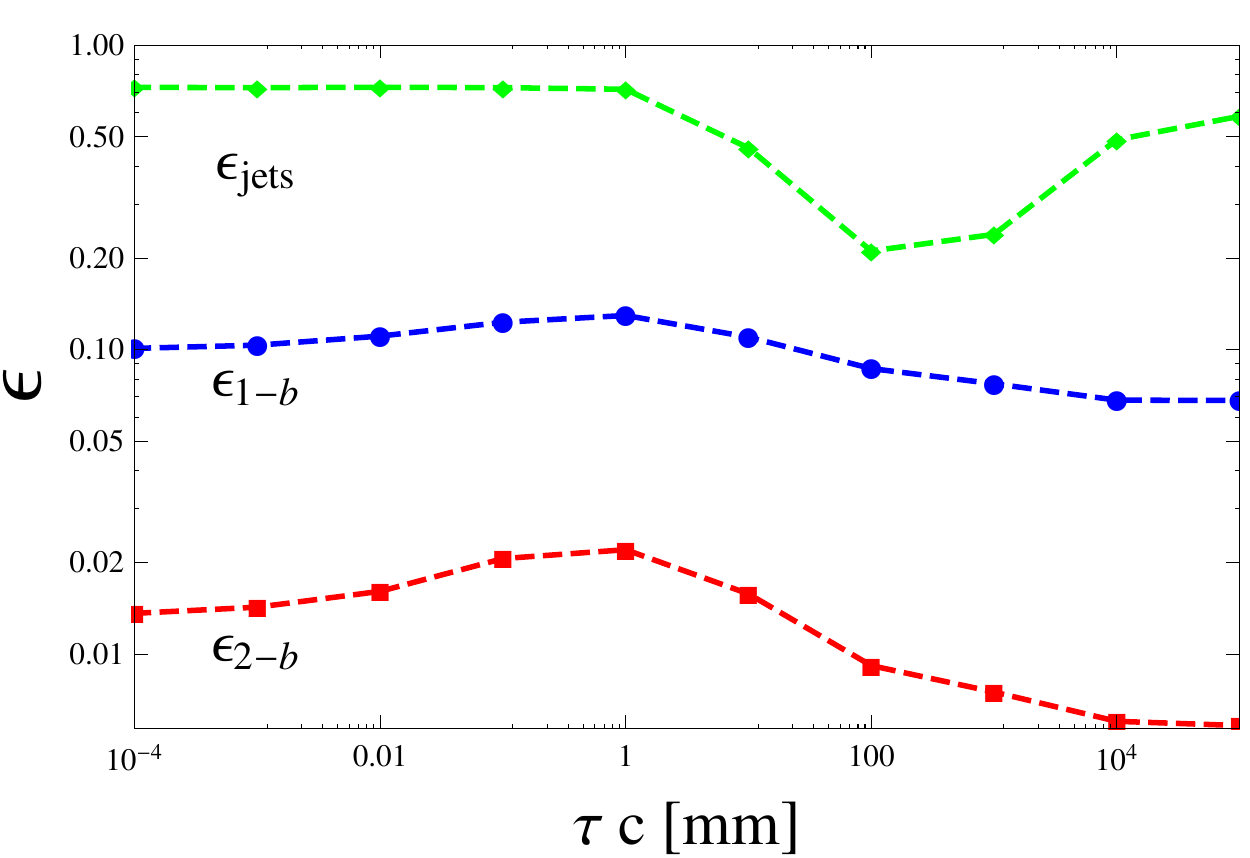}~
\caption{ \small Efficiencies for event reconstruction. The green diamonds denote the acceptance rate that all light jets in an event contain at least one well reconstructed charged track (transverse impact parameter below $\sim2$ mm) as a function of the colored scalar lifetime. The blue circles and red squares denote the $1$ and $2$ $b$-tagging efficiencies for a heavy flavor quark decaying at larger displaced vertices than those predicted by the SM.  }\label{fig:Fig2}
 \end{figure*}
Collider signatures from long-lived particles are relatively free from SM backgrounds. However, conventional searches are limited to impact parameters below ${\cal O}\left(1\right)$ mm. This setup is mainly motivated to probe decays from heavy flavor quarks and $\tau$-leptons. In addition, most searches incorporate track quality cuts with requirements on the impact parameters of charged tracks~\cite{Chatrchyan:2011ds,Aad:2013gja} and use impact parameter-base tagging~\cite{Chatrchyan:2012jua,ATLASbTag,Chatrchyan:2012zz,TheATLAScollaboration:2013wha} for heavy flavor quarks and $\tau$'s. The latter poses a problem for long-lived particles decaying to $b$-jets, since this scenario will lead to displaced vertices from $B$ decays with impact parameters above $\sim 1$ mm; thus leading to a large suppression on the $b$-tagging efficiency. In order to study the properties of jets as a function of the average lifetime of a decaying colored scalar scalar we use Delphes~\cite{delphes} for jet reconstruction and tracking information. We simulate the detector response for events for a slightly compressed scenario of the model discussed in the previous section using $m_{\phi}=150$ GeV and $m_{X}=20$ GeV. In Figure~\ref{fig:Fig2} we show in green the efficiency that all jets in the events contain at least one well reconstructed charged track as a function of the particle's lifetime. We use the ATLAS guidelines for tracks~\cite{Andreazza:2009zz} that place standard quality cuts on the transverse impact parameter, $|d_{0}|<2$ mm, and the difference between the longitudinal impact parameter and the position of the primary vertex, $|z_{0}-z_{v}|<10$ mm. We observe a decrease in the efficiency for decay lengths above $1$ mm and an increase above $1$ m. Above $10$ mm, a jet will still contain a number of tracks but many will have impact parameters not consistent with the quality criteria. At and above $1$ m, the long-lived particle decays within the calorimeter. Most tracks found inside this jet will mostly come from activity in the inner detector; these jets will have a higher probability of satisfying the quality criteria.

The $b$-tagging efficiency is depicted by the blue line within Figure~\ref{fig:Fig2} while the two $b$-tagging efficiency is depicted by the red line. In Delphes, the default $b$-tagging efficiency is of $40\%$ for $b$-jets with $p_{T}>20$ GeV and $|\eta|<2.5$. We are using a compressed spectrum, where the $b$-tagging efficiency for $b$'s originating from a prompt decay is of $20\%$. In addition, Delphes does not implement any quality criteria on the impact parameters on the tracks that arise from the decay of a $B$-hadron. We apply the quality criteria in~\cite{Rizzi:2006ms,CMS:2012rta} and observe that for $b$'s from prompt decays the efficiency is on the order of $10\%$ with a slight decrease on the $b$-tagging efficiency above $1$ mm. The loss in efficiency is even greater for the two $b$-tagging efficiency. Therefore, to probe long-lived particle decays, especially within the inner detector, analyses must incorporate dedicated triggers and beyond the accustomed particle reconstruction algorithms. The latter should have the potential to pin down the origin of many particles (tracks) to one single displaced vertex~\cite{CMS:2014wda,TheATLAScollaboration:2013yia,ATLASnew}.

\begin{figure*}[ht]
\centering
\includegraphics[width=10.0cm]{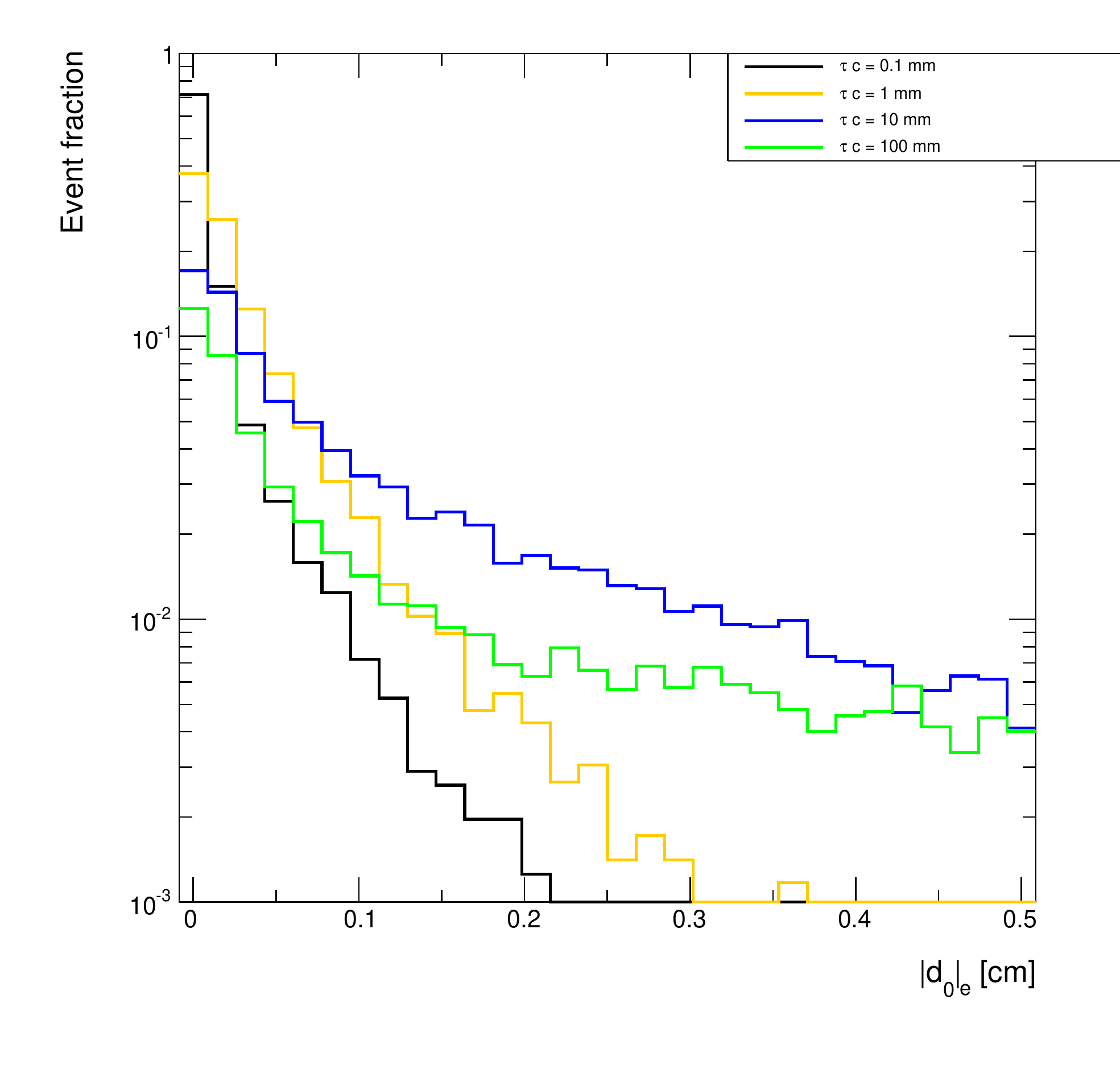}~
\caption{ \small Fraction of events as a function of the electron's transverse impact parameter for colored scalars with decay lengths in the rage $0.1-100$ mm. }\label{fig:Fig3}
 \end{figure*}
Our simplified model has new colored degrees of freedom that can be pair produced. The colored scalars decay to a dark fermion that is a source of missing transverse energy and to a top quark, and depending on the spectrum and coupling strength, 2-, 3-, and 4- body decay modes may be open. The latter is important in the region where the scalar and dark fermion have a small mass gap and the decay leads to soft jets or leptons that may escape detection since they may not be fully reconstructed. Therefore, most of the SUSY searches with zero leptons will have a potential reach to long-lived colored scalars that decay with lifetimes above $~0.1$ mm$/c$. For this class of particles the decay products, jets, will have impact parameters within the thresholds used in prompt SUSY searches. However, a significant fraction of leptons will not be identified, since tracks with impact parameters above $\sim1$ mm are not considered in conventional collider searches. This can be seen qualitatively in Figure~\ref{fig:Fig3}, where we show the fraction of events as a function of the electron's impact parameter for decay lengths in the range between $0.1-100$ mm.

\begin{figure*}[ht]
\centering
\includegraphics[width=10.0cm]{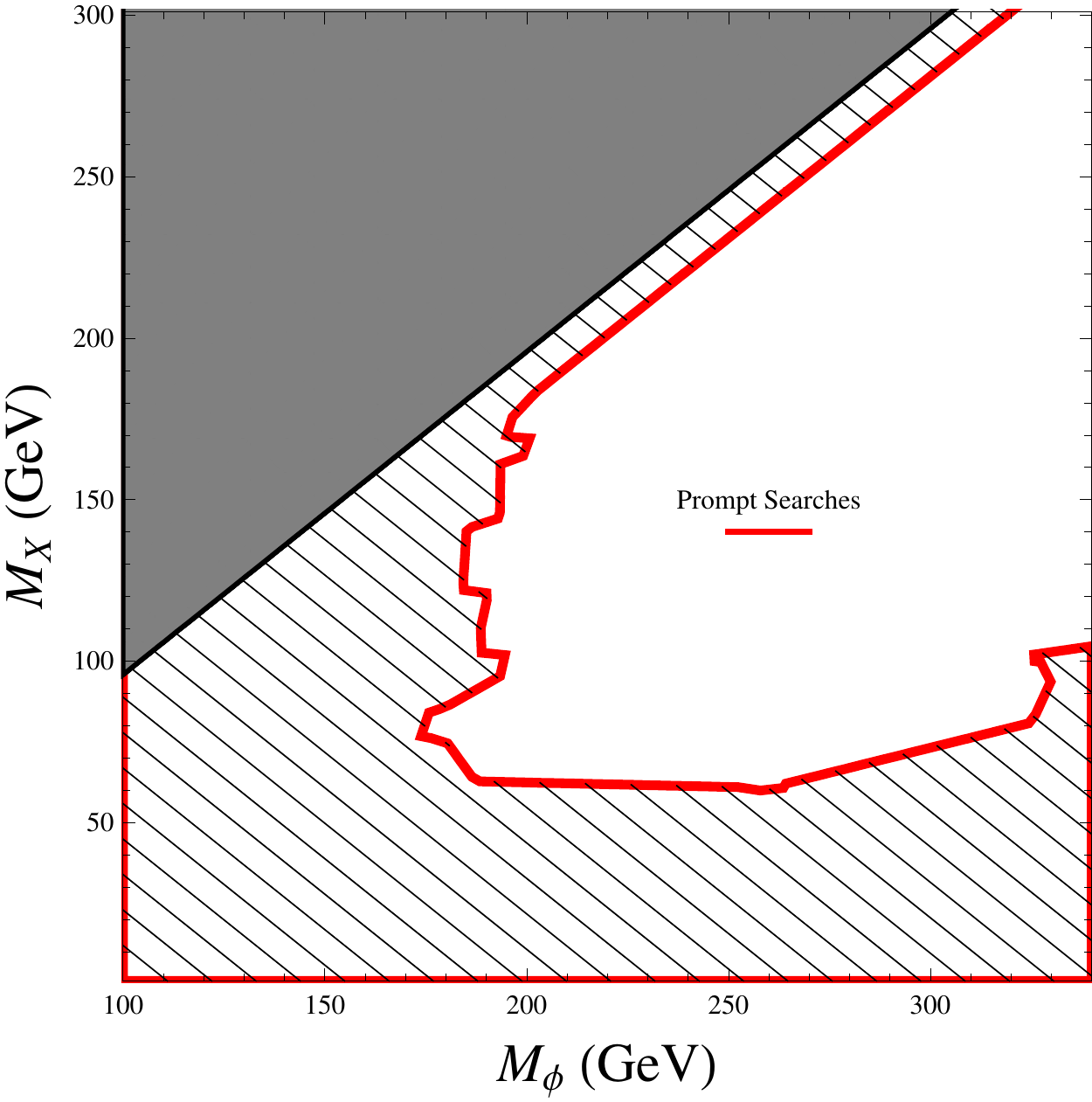}~
\caption{ \small  Excluded regions in the $m_{\phi}-m_{X}$ plane with the prompt searches included in this sections at $8$ TeV center of mass energies. We assume the exclusion applies to lifetimes between $0.1$ and $10$ mm$/c$ since no additional suppression on the signal acceptance is expected from the displaced jets in the event. }\label{fig:Fig4}
 \end{figure*}

In order to analyze the reach of prompt searches at the $8$ TeV LHC to our simplified scenario, we use the analyses validated for phenomenological use by the CheckMATE collaboration~\cite{checkmate}. To achieve the largest sensitivity, we analyze particle lifetimes in the range $0.1-10$ mm$/c$ using the following collider searches by the ATLAS and CMS collaborations:
\begin{itemize}
\item
$1~\text{lepton}+4~\text{jets}+\slashed{E}_{T}$~\cite{ATLAS:2012tna}
\item
$\text{Monojet search}+\slashed{E}_{T}$~\cite{ATLAS:2012zim}
\item
$\text{Stops with monojets and charms}+\slashed{E}_{T}$~\cite{Aad:2014nra}
\item
$0~\text{lepton}+6~(2b)\text{-jets}+\slashed{E}_{T}$\cite{ATLAS:2013cma}
\item
$2-6~\text{jets}+\slashed{E}_{T}$\cite{TheATLAScollaboration:2013fha}
\item
$2~\text{leptons}+~\text{jets}+\slashed{E}_{T}~(razor)$\cite{TheATLAScollaboration:2013via}
\item
$\text{At least 2 jets}+b~\text{jet multiplicity}+\slashed{E}_{T}(\alpha_{T})$\cite{Chatrchyan:2013lya}

\end{itemize}
\begin{figure*}[ht]
\centering
\includegraphics[width=10.0cm]{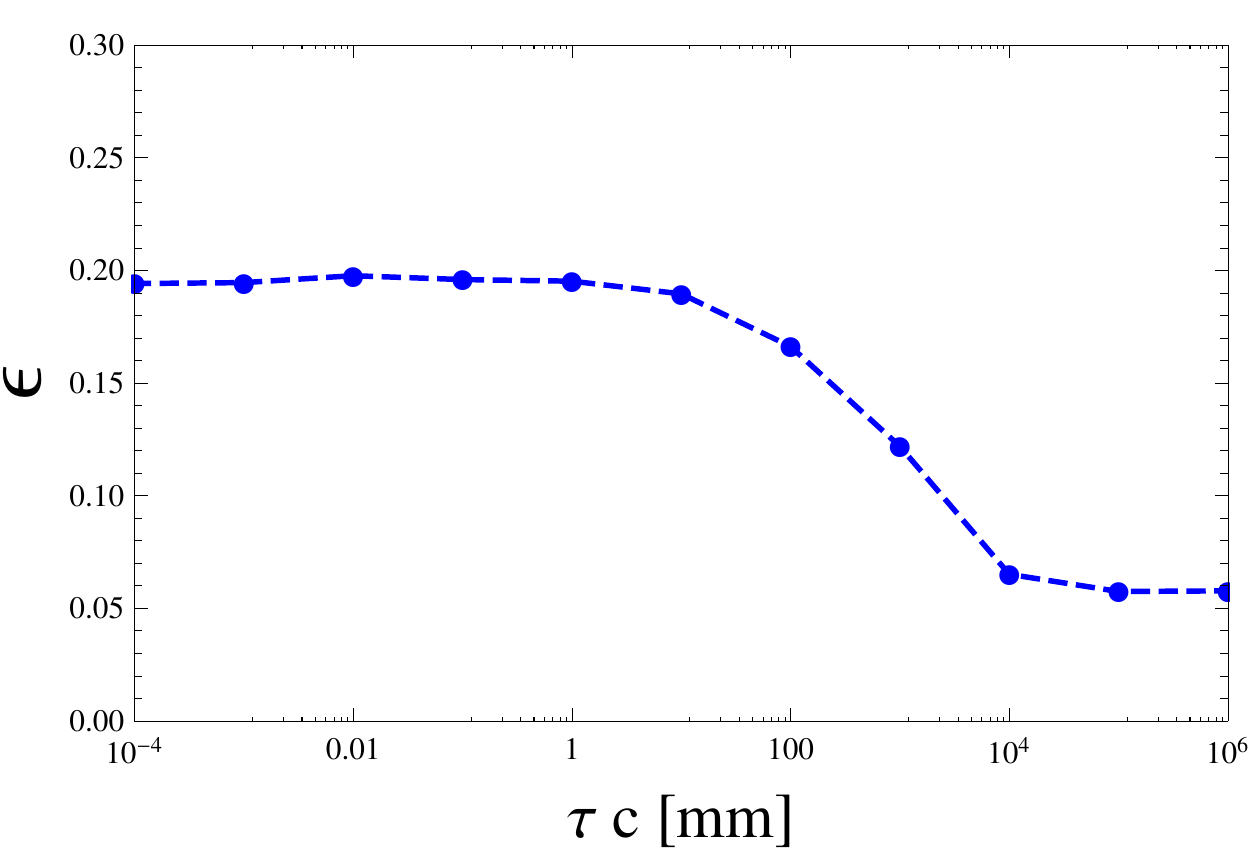}~
\caption{ \small  Efficiencies for event reconstruction. The blue circles denote the acceptance rate that an event contains between one and two jets with $p_{T}>100$ GeV, a charge $p_{T}$ fraction, $f_{ch}>0.02$, and ($f_{ch}>0.05$ or $f_{em}<0.9$). }\label{fig:Fig5}
 \end{figure*}
In Figure~\ref{fig:Fig4} we show the region of parameter space in the $m_{\phi}-m_{X}$ plane excluded with the ATLAS and CMS analyses described above for a colored scalar with a $0.1$ mm decay length. In particular, we observe that the CMS analysis using the $\alpha_{T}$ variable~\cite{Chatrchyan:2013lya} yields the largest exclusion in the non-degenerate region while the monojet~\cite{Aad:2014nra} analysis constraints the degenerate window, where most of the signal is due to a hight-$p_{T}$ jet from initial state radiation. The monojet constraint is very insensitive to the particle's lifetime. However, the region excluded by searches incorporating large number of jets can be suppressed as the particle's lifetime increases. Recently, the ATLAS collaboration has re-casted a number of prompt searches to constrain a long-lived gluino~\cite{stGluinos}. In particular, these searches require large number of jets and no leptons. For example, the ATLAS search for squarks and gluinos~\cite{1405.7875} requires a leading jet with $p_{T}$ above $130$ GeV in all of its signal regions, and a second jet with $p_{T}>60$. GeV. Many ATLAS analyses reject events containing a jet with $p_{T}>100$ GeV and $|\eta|<2$ if the charged $p_{T}$ fraction satisfies, $f_{ch}<0.02$ or both $f_{ch}<0.05$ and an electromagnetic fraction, $f_{em}$, above 0.9 to reject cosmics or detector malfunctions~\cite{1308.2631,Aad:2011he}. In addition, jets fail reconstruction if they appear to originate from additional collisions, that is, jets with $p_{T}>50$ GeV and $|\eta|<2.5$ are required to have at least one well reconstructed track as described above. In Figure~\ref{fig:Fig5} we show the acceptance rate obtained after demanding between one and two jets with $p_{T}>100$ GeV and satisfying the above quality criteria. The acceptance rate is flat for lifetimes below $\sim 100$ mm$/c$. If the highest efficiency (for small lifetimes) corresponds to the efficiency of current prompt searches, then prompt searches will loose sensitivity for lifetimes above $100$ mm$/c$. In particular, for decay lengths between $100$ mm and $10$ m, sensitivity of prompt searches would be suppressed by roughly $10-60\%$ respectively. This region of parameter space is then very difficult to probe with prompt searches as well as displaced vertex searches since in the latter, a full reconstruction of the secondary vertex is required, which can only happen at displacements from the interaction point below $600$ mm. Above $\sim1-10$ m, the long-lived particle decays within the calorimeter and most tracks found within the jet will most likely come from activity in the inner detector. This is depicted by the flat efficiency found above $1-10$ m. We conclude that in our particular simplified scenario, prompt searches with lepton vetoes are incredibly effective for lifetimes below $10$ mm$/c$. In light of this result, which falls short of an in depth analysis of a collider detector, we encourage the collaborations to continue their studies regarding the applicability of prompt searches to lifetimes above $~0.1$ mm$/c$ in order to better pinpoint the region where more specialized triggers must be used; to both address interesting physics and the issue of pile-up at a stronger hadron collider.

In the following section we analyze a long-lived search by the CMS collaboration that can further probe the parameter space of this model, in particular, regions where displaced leptons from the decays of colored scalar decays appear with impact parameters between $0.01$ and $2$ cm.

\subsection{Long-lived searches at LHC 8 TeV}\label{subsec:LLS8}
\begin{figure*}[ht]
\centering
\includegraphics[width=7.0cm]{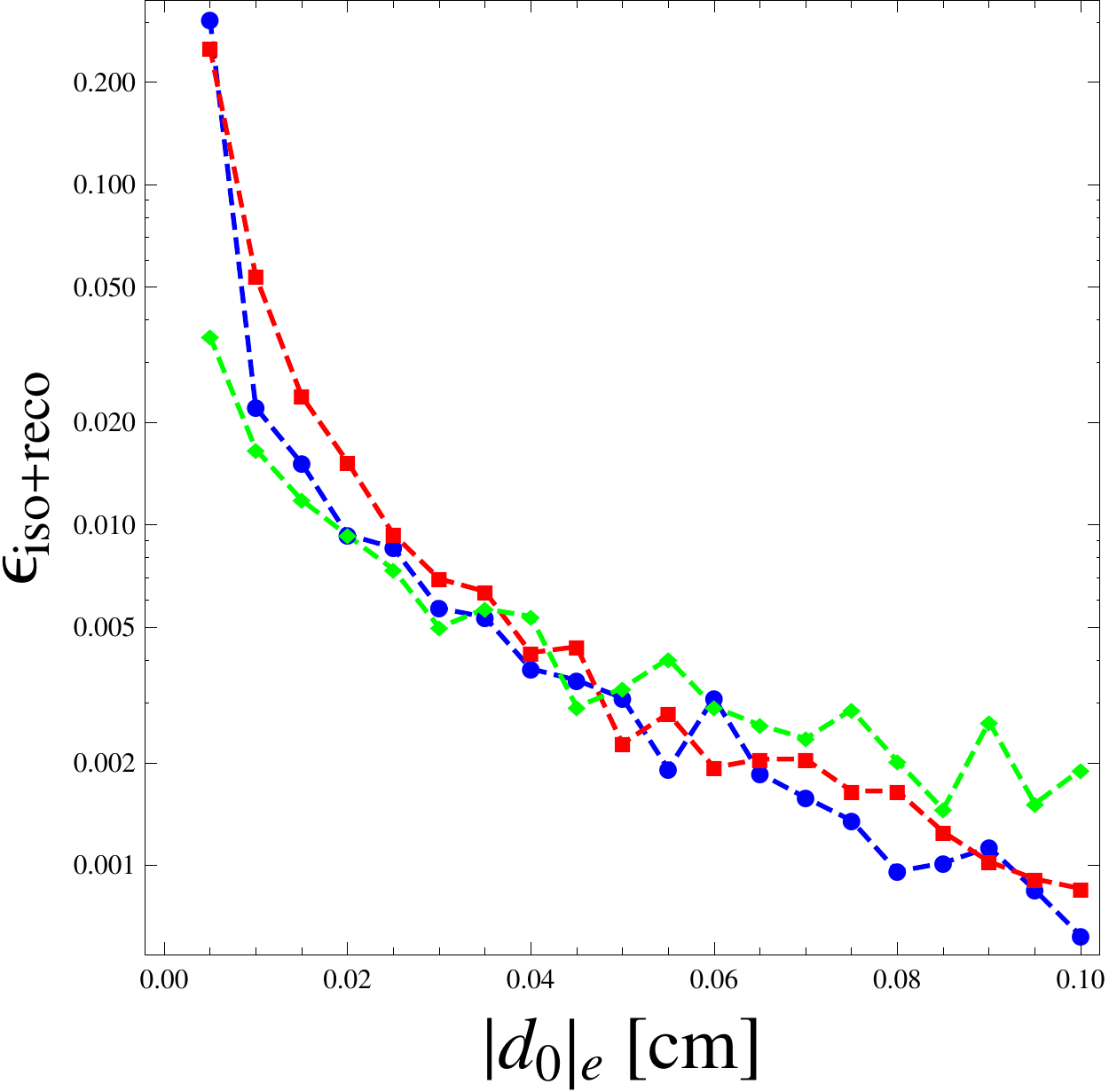}
\caption{ \small Electron isolation and reconstruction efficiency as a function of the transverse impact parameter. The blue dashed line corresponds to a long-lived decaying particle with a mean lifetime of $\tau=0.0001$ mm$/c$ while the red, and green lines correspond to mean lifetimes of $0.1$ and $100$ mm$/c$ respectively.}\label{fig:iso}
 \end{figure*}
 In this section we focus on an analysis by the CMS collaboration that searches for long-lived charged massive particles through their decays into leptons with $19.5$ fb$^{-1}$ of data at $8$ TeV center of mass energies~\cite{Khachatryan:2014mea}. In particular, this search is sensitive to pair produced long-lived stops that decay to a $b$-quark and a lepton through an $R$-parity violating vertex~\cite{Graham:2012th}. The search is designed to be as model independent as possible and focuses exclusively on a displaced isolated dilepton signature. In particular, the final state consists of exactly one electron and one muon, oppositely charged, and isolated within a cone $\Delta R=0.3,04$ respectively. The leptons are required to have a $p_{T}> 25$ GeV and lie within the detector acceptance of $|\eta|<2.5$. The two leptons are required to be separated in the $\eta -\phi$ plane by more than $0.5$. Events are rejected if a jet with $p_{T}>10$ lies within $\Delta R<0.5$ of either selected lepton.
 
 The main background leading to a displaced lepton in association with jets is from $Z$ production with decay into a $\tau^{+}\tau^{-}$ final state. QCD multijet and $t\bar{t}$ production are also dominant sources of background since a displaced lepton may appear from the misidentification of a jet and from the decay of a heavy flavor jet. $B$ and $D$ meson decays lead to displaced leptons with a mean displacement of $\sim 500~\mu$m while leptons from $\tau$ decays have a mean lifetime of $\sim87~\mu$m. Other sources of background come from single top production, diboson, and $W+$ jets production. In order to validate our analysis, we simulate the $Zj(Z\to\tau^{+}\tau^{-})$ and the $t\bar{t}$ backgrounds, and use the data driven QCD multijet background results quoted by the CMS collaboration. A Monte Carlo approach for the QCD component will require a very large sample to properly handle the multijet background. We use MadGraph 5~\cite{Alwall:2011uj} in order to simulate the backgrounds and the pair production of a colored scalar with model files generated with FeynRules~\cite{Feynr}. The parton showering and hadronization are carried out by Pythia~\cite{Sjostrand:2006za}, as well as the decays of the colored scalar which are implemented by a decay table that takes into account the 2-,3-, and 4-body decay branching fractions. In our analysis we use the Delphes~\cite{delphes} fast detector simulator to reconstruct jets with the anti-$k_{T}$ algorithm using a parameter size of $\Delta R=0.5$, and to isolate and reconstruct tracks. The lepton identification is performed by matching a track to a truth-level lepton. Furthermore, in the simulation of the backgrounds we further match the leptons to the decay of a $\tau$ or a heavy flavor jet. The electron reconstruction efficiencies are implemented using data from the CMS collaboration on electron performance~\cite{CMS:2013hoa} and we implement a muon reconstruction efficiency of $95\%$. The isolation, $I_{l}$, criteria is calculated by summing the $p_{T}$ of all tracks within a cone of size $\Delta R=0.3$ and $0.4$ around the electron and muon respectively. We require that 
\begin{equation}
 I_{l}=\frac{\sum_{i\ne l}p_{T,i}}{p_{T,l}} < 0.10~(l=\text{electron}),~0.12~(l=\text{muon})
 \end{equation}

Events passing the preselection criteria described above are further categorized by their transverse impact parameter. The transverse impact parameter is calculated using a track's coordinate of closest approach to the primary vertex in the transverse plane, $x_{d},y_{d}$, and the track's transverse momentum:
\begin{equation}
|d_{0}|=\frac{|x^{track}_{d}p^{track}_{y}-y^{track}_{d}p^{track}_{x}|}{p^{track}_{T}}\label{eq:IP}
\end{equation}
 \begin{figure*}[ht]
\centering
\includegraphics[width=7.0cm]{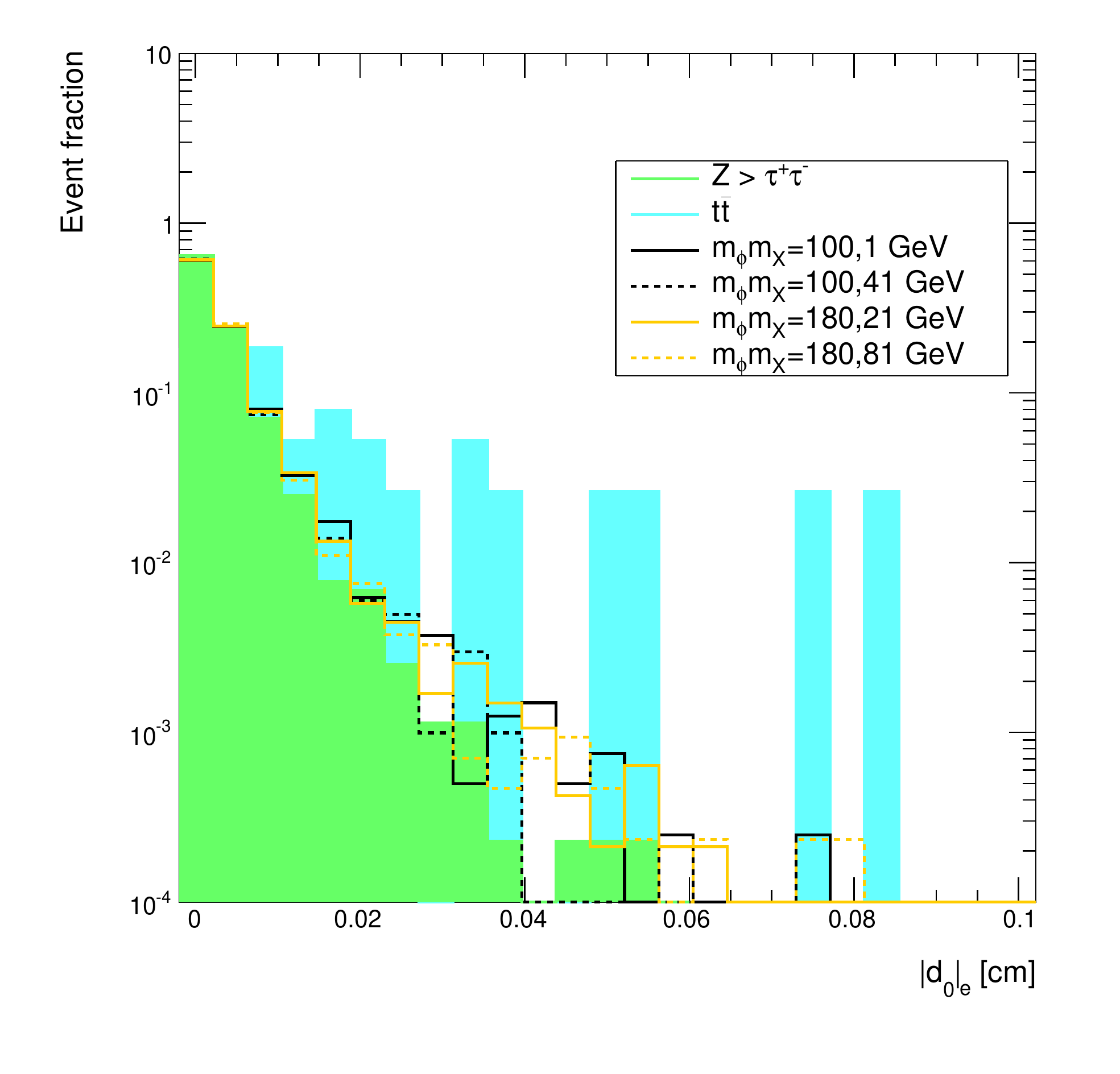}~
\includegraphics[width=7.0cm]{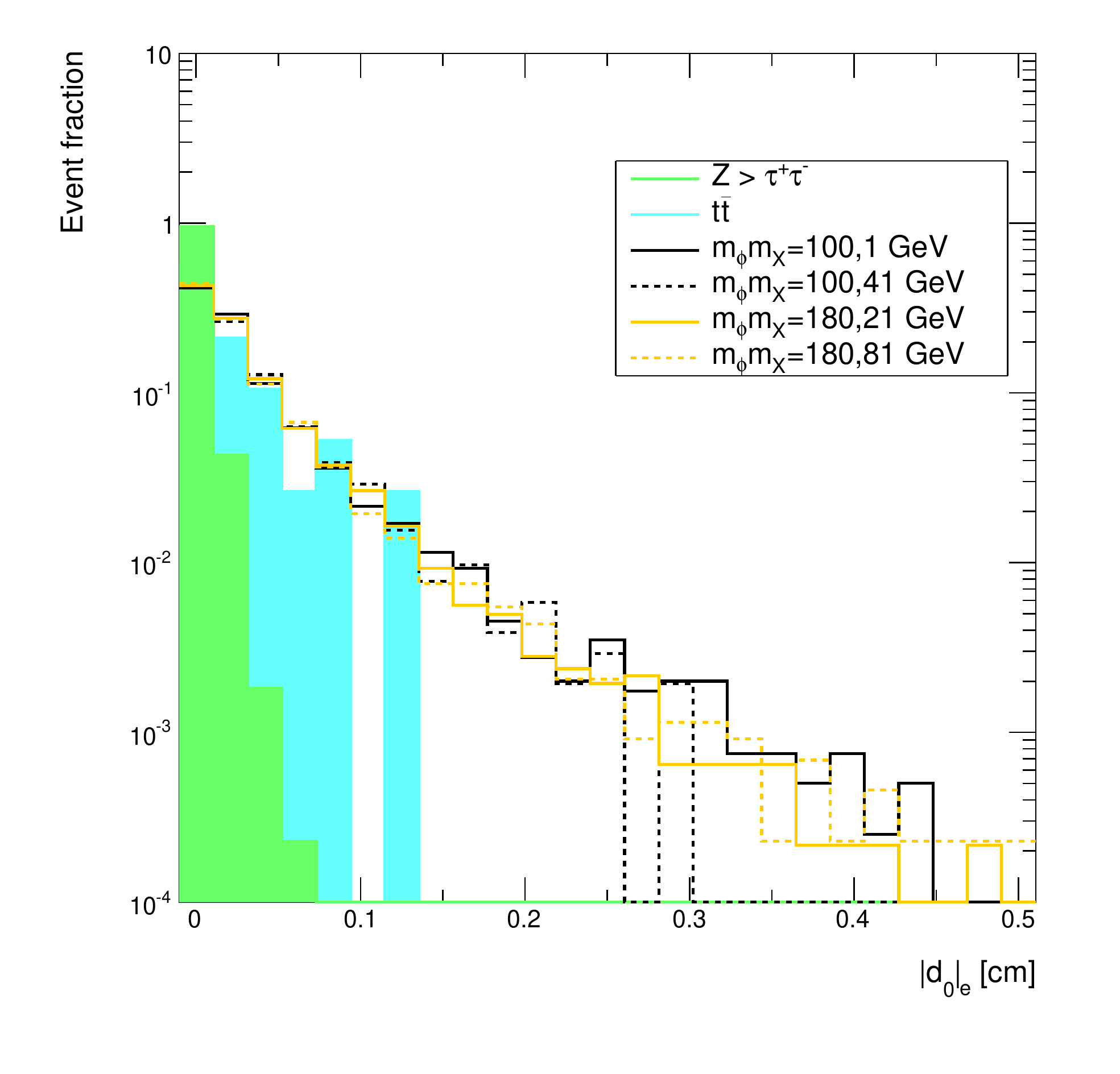}~\\
\includegraphics[width=7.0cm]{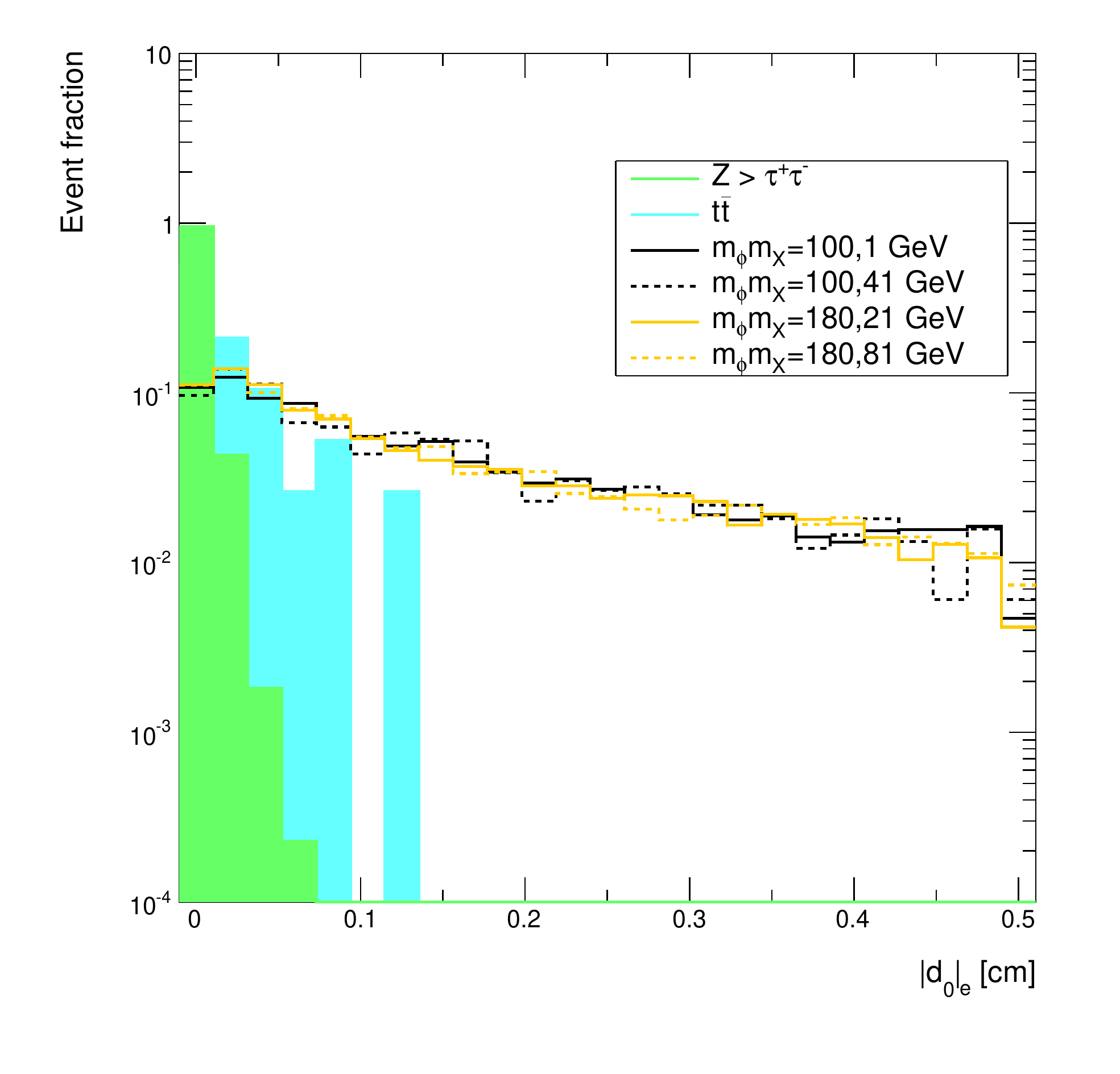}
\caption{\small Fraction of events as a function of the electron's transverse impact parameter, $|d_{0}|_{e}$. The top left figure corresponds to a colored electroweak-singlet scalar decay length of $0.1$ mm while the top right figure corresponds to $1$ mm. The figure at the bottom corresponds to a decay length of $10$ mm.  }\label{fig:d0E8}
 \end{figure*}
\begin{figure*}[ht]
\centering
\includegraphics[width=7.0cm]{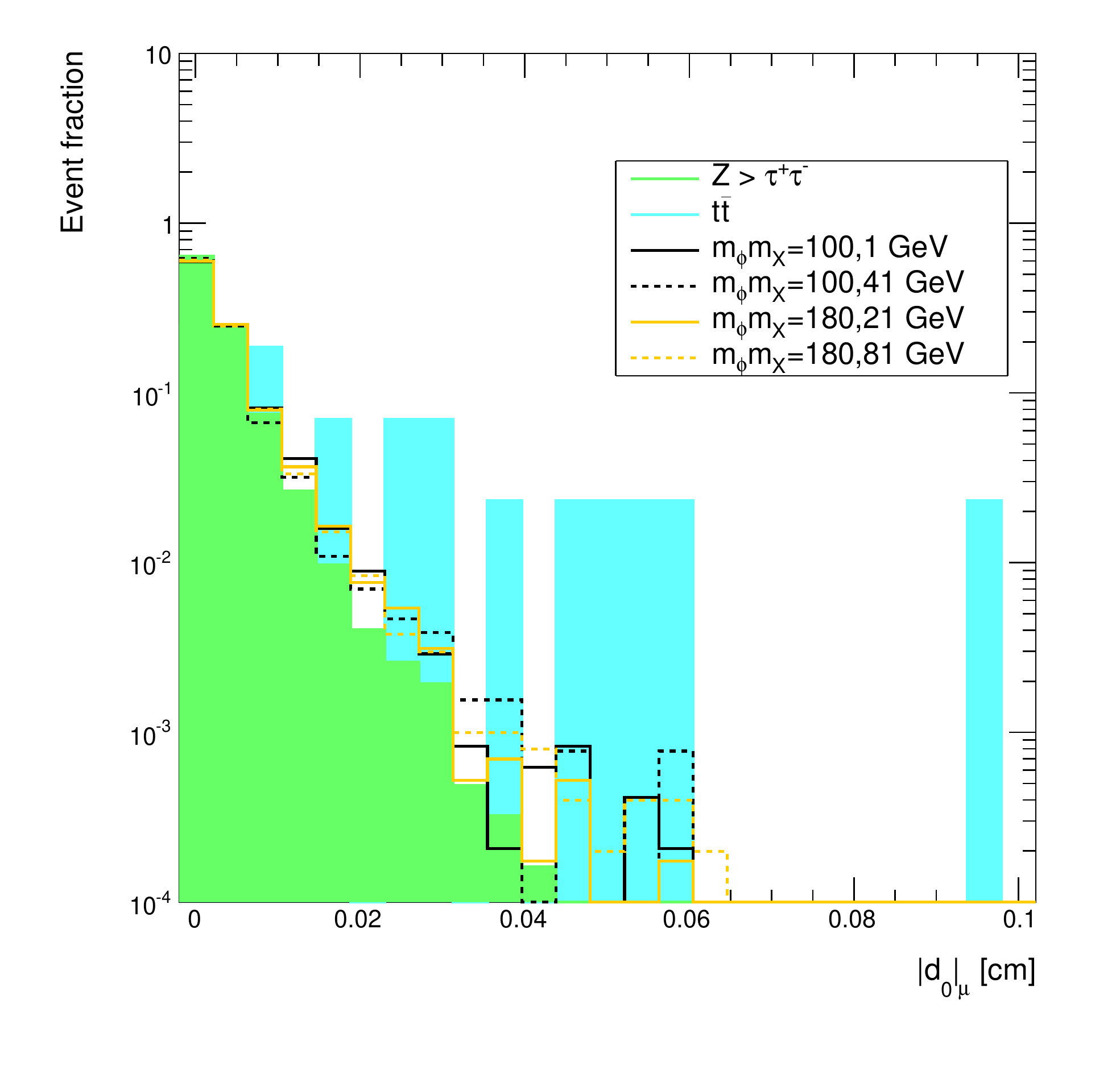}~
\includegraphics[width=7.0cm]{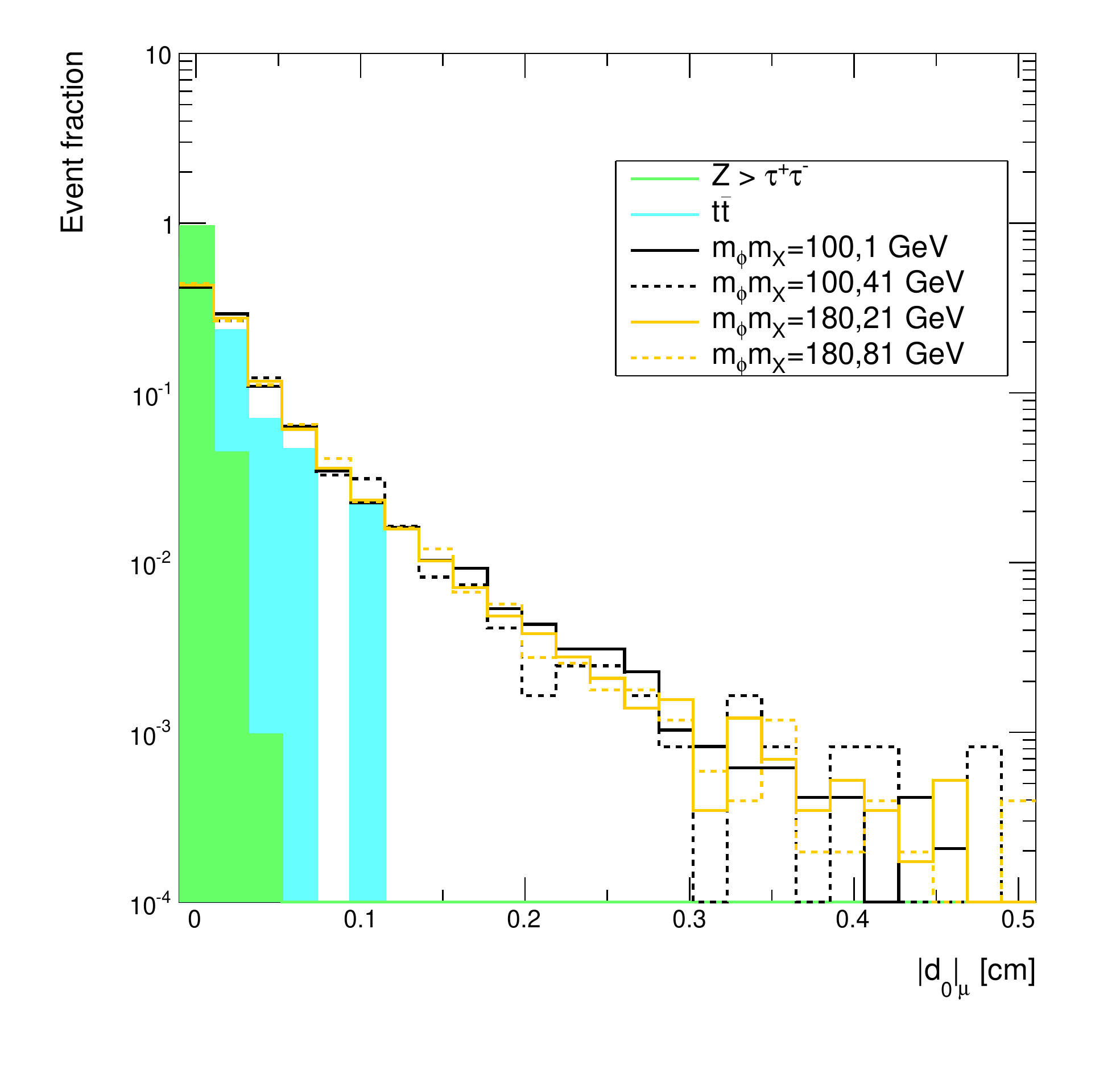}~\\
\includegraphics[width=7.0cm]{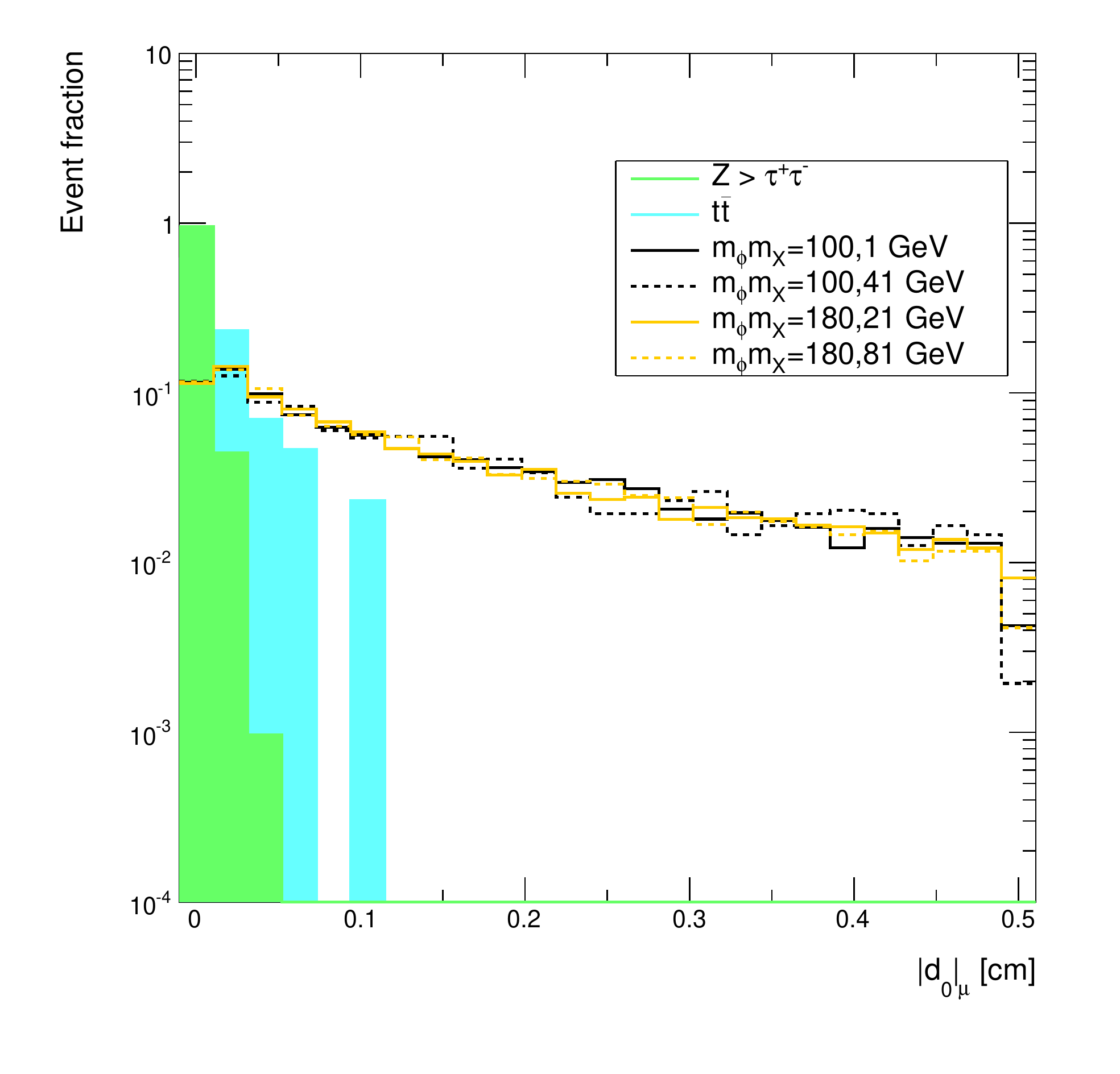}
\caption{\small Fraction of events as a function of the muon's transverse impact parameter, $|d_{0}|_{\mu}$. The top left figure corresponds to a colored electroweak-singlet scalar decay length of $0.1$ mm while the top right figure corresponds to $1$ mm. The figure at the bottom corresponds to a decay length of $10$ mm. }\label{fig:d0M8}
 \end{figure*}
In order to exclude contamination from promptly decaying particles, both leptons are required to have an impact parameter, $|d_{0}|$ above 100 $\mu$m. This requirement eliminates almost completely subdominant sources of backgrounds that yield an electron or a muon from the decay of a $W$, $Z$ and a top quark. Lastly, the analysis is constrained within a region where $|d_{0}|< 2.$ cm. In order to properly validate our analysis and to accurately constrain the parameter space of the model introduced in this work we analyze the dependence of the reconstruction and isolation efficiencies as a function of the impact parameter in a pure sample of displaced leptons arising from the decay of a colored scalar with mass $m_{\phi}=150$ GeV. The results are shown in Figure~\ref{fig:iso}. The blue dashed line corresponds to a long-lived decaying particle with a mean lifetime of $\tau=0.0001$ mm while the red and green dashed lines correspond to mean lifetimes of $0.1$ and $100$ mm$/c$ respectively. The plot shows a clear relationship between a particle's lifetime and the impact parameter of the decay product and it is consistent with what we observe in Figure~\ref{fig:Fig3}. We observe in Figure~\ref{fig:iso} that even though leptons from decaying particles with lifetimes of $100$ mm/$c$ are likely to appear with very small impact parameters, the efficiency for proper isolation and reconstruction is smaller than leptons arising from decays of shorter lived particles at these small values of impact parameters.
 
We simulate the signal pertaining to the model described in the previous section for three colored electroweak-singlet scalar decay lengths: $0.1, 1$ and $10$ mm in the range of masses $m_{\phi}<350$ GeV and $m_{X}<300$ GeV as well as the $Zj(Z\to\tau^{+}\tau^{-})$ and $t\bar{t}$ backgrounds applying the preliminary cuts described above but only demanding that there is either one electron or one muon. The results are depicted in Figures~\ref{fig:d0E8}-\ref{fig:d0M8} where we show the transverse impact parameters for the electron and muon respectively. In the figures the black solid line corresponds to $m_{\phi}=100$ GeV and $m_{X}=1$ GeV. Here, the main decay mode of $\phi$ is dominated by the three-body channel into $X$, a $b$-jet and a $W$ boson. The dashed back line, solid and dashed yellow lines correspond to values of $m_{\phi} (m_{X})=100~(41),180~(21),180~(81)$ GeV respectively. The leptons in the $Zj$ sample (green histogram) arise from leptonic decays of the $\tau$ while leptons in the $t\bar{t}$ sample (blue histogram plotted behind the $Zj$ background) arise mainly from semileptonic $b$-meson decays. These leptons arise from the decay of an off-shell $W$ boson from the dominant decay mode $b\to cW^{*}$. Therefore the region of impact parameters in the range $0.02 - 0.5$ cm allows us to directly probe this model with the existing data from the CMS collaboration.
\begin{table}[ht]\centering
 \tabcolsep 2.2 pt
\small
\begin{tabular}{|c|c|c|c|c| }
\hline
SM background &  $SR_{1}$ & $SR_{2}$ & $SR_{3}$  \\
\hline
\hline
Total Expected Background &  $18.0\pm0.5\pm3.8$ & $1.01\pm0.06\pm0.30$ & $0.051\pm0.015\pm0.010$  \\ 
\hline
Observed Events & $19$ & $0$ & $0$  \\ 
\hline
$95\%$ CL$_{s}$ &  $11$ & $3$ & $3$ \\
\hline
\hline
$p~p\to \tilde{t}_{1}\tilde{t}^{*}_{1}$ (RPV)  & \multicolumn{3}{c|}{} \\
\hline
$M=500$ GeV, $\left<c\tau\right>=$ $1$ mm & $19.7~(30.1\pm0.7\pm5.3)$ & $5.8~(6.54\pm0.34\pm1.16)$ & $1.03~(1.34\pm0.15\pm0.24)$ \\
\hline
$M=500$ GeV, $\left<c\tau\right>=$ $10$ mm & $28.7~(35.3\pm0.8\pm6.2)$ & $26.4~(30.3\pm0.7\pm5.3)$ & $49.6~(51.3\pm1.0\pm9.0)$ \\
\hline
$M=500$ GeV, $\left<c\tau\right>=$ $100$ mm & $5.2~(4.73\pm0.30\pm0.83)$ & $6.2~(5.57\pm0.32\pm0.98)$ & $39.2~(26.3\pm0.7\pm4.6)$ \\
\hline

\end{tabular}
\caption{\small Number of expected and observed events given by the CMS analysis on displaced SUSY~\cite{Khachatryan:2014mea}. The third row corresponds to the number of excluded signal events at $95\%$ CL$_{s}$. In addition, we have re-casted the model analyzed in the CMS search and show the number of events within each signal region for the three benchmark points in our simulation and in parentheses those of CMS.} \label{tab:eventsEx}
\end{table}

In order to constrain the parameter space of our simplified model we implement the CMS analysis which is performed in three signal regions parametrized by the transverse impact parameter: The most exclusive region requires that both the electron and muon have an impact parameter above $0.1$ cm. This region is labeled $SR_{3}$. The intermediate region, $SR_{2}$ is populated with events not in $SR_{3}$ but with both leptons above $0.05$ cm. $SR_{1}$ is populated with events failing the $SR_{2}$ criteria but with both leptons above $0.02$ cm. We carry out a channel by channel exclusion using a $95\%$ CL number of excluded signal events. In Table~\ref{tab:eventsEx} we show the number of observed and expected events calculated in the CMS analysis as well as the number of events to make an exclusion at $95\%$ confidence level. We fully recast the $Z\to\tau\tau$ background in the window of impact parameters $0.01<|d_{0}|<2$ cm after implementing all preliminary cuts and obtain $82$ events v.s. the $98$ obtained by CMS. However to make the exclusion we make use of the experimental results due to our inability to fully recast the $t\bar{t}$ and QCD backgrounds. In addition, we simulate the RPV decay mode $\tilde{t}\to b~l$ analyzed by the CMS collaboration and show that our results are consistent with the experiment.

\begin{figure*}[ht]
\centering
\includegraphics[width=7.0cm]{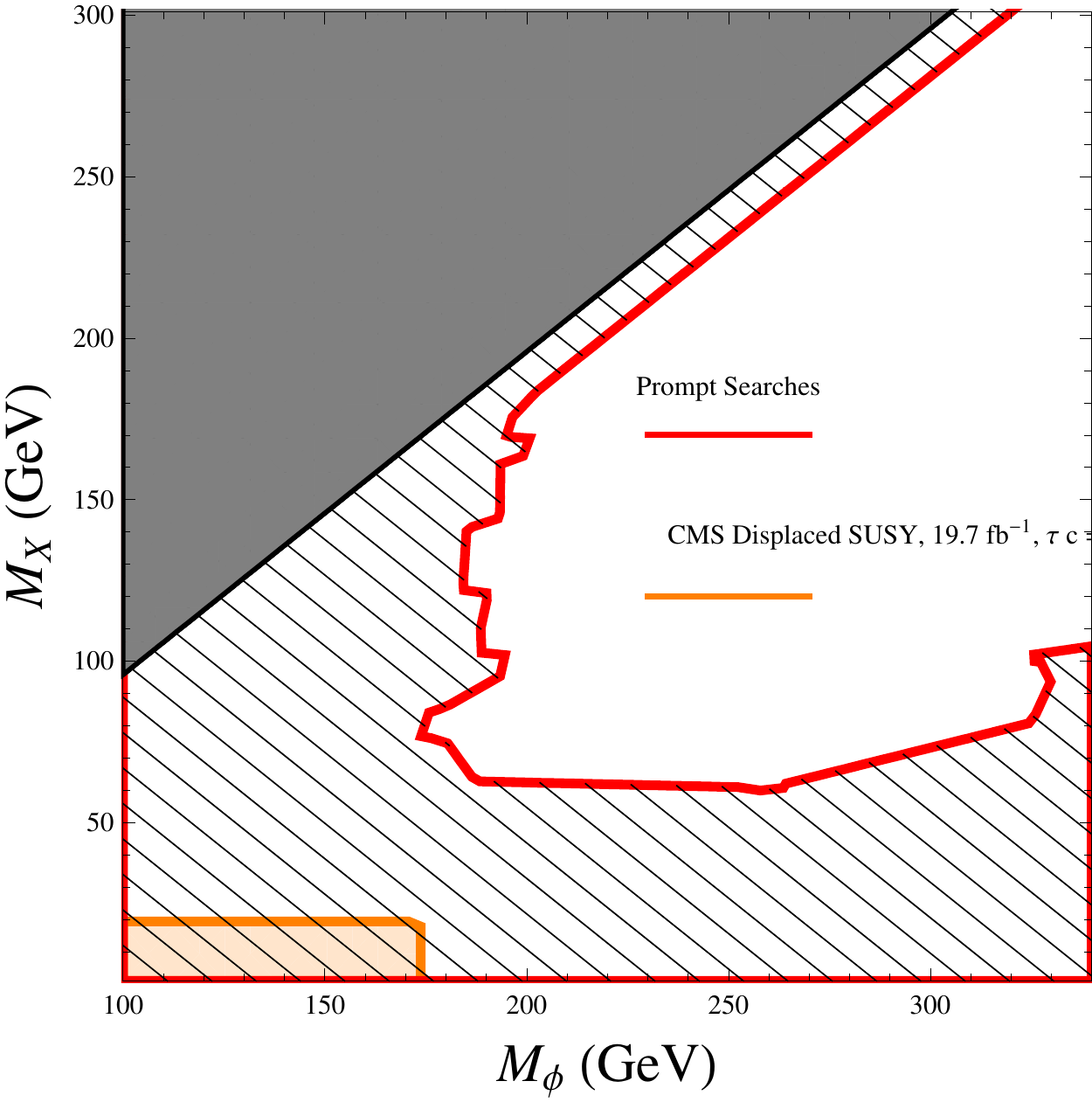} \\
\includegraphics[width=7.0cm]{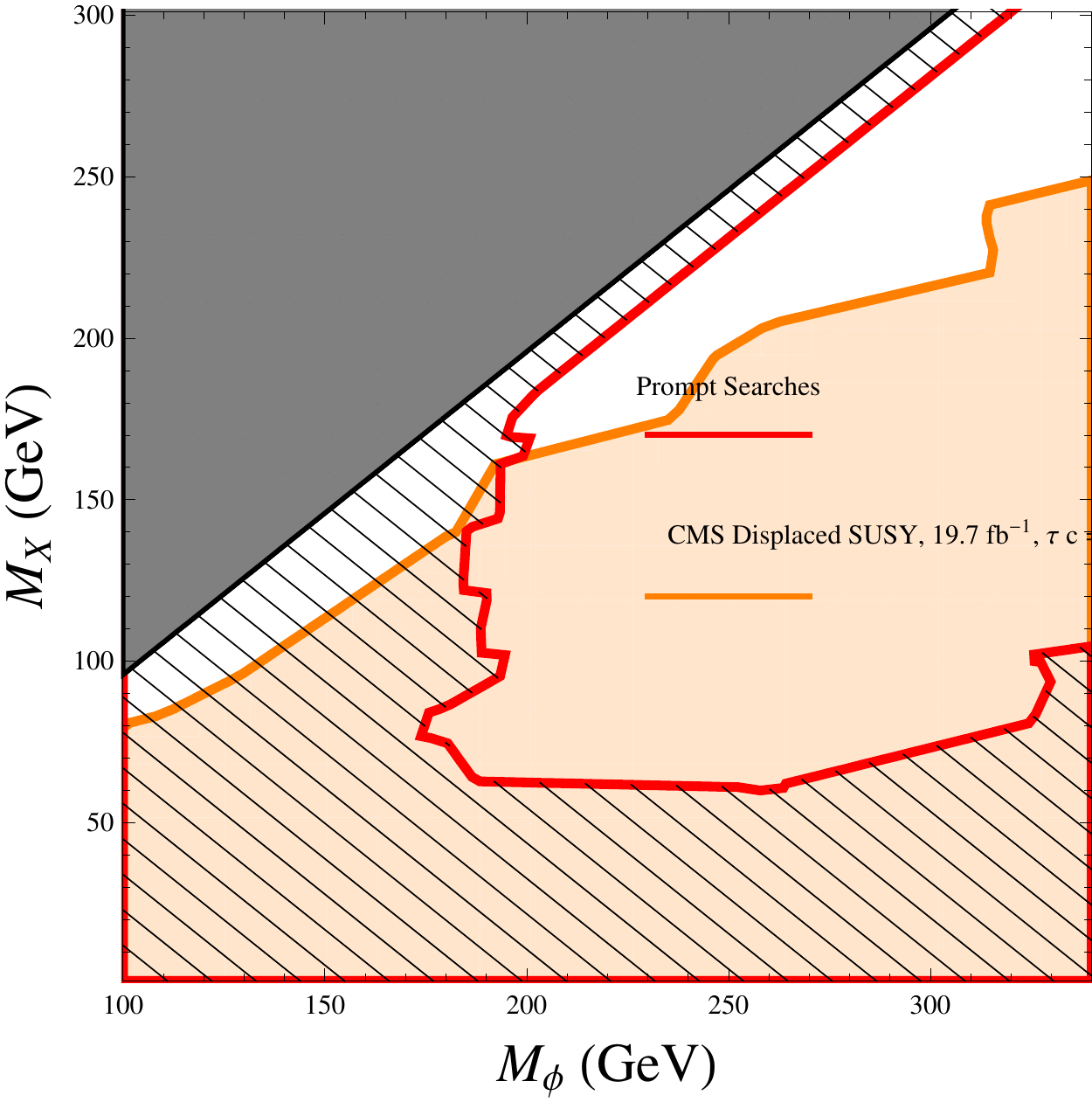}~
\includegraphics[width=7.0cm]{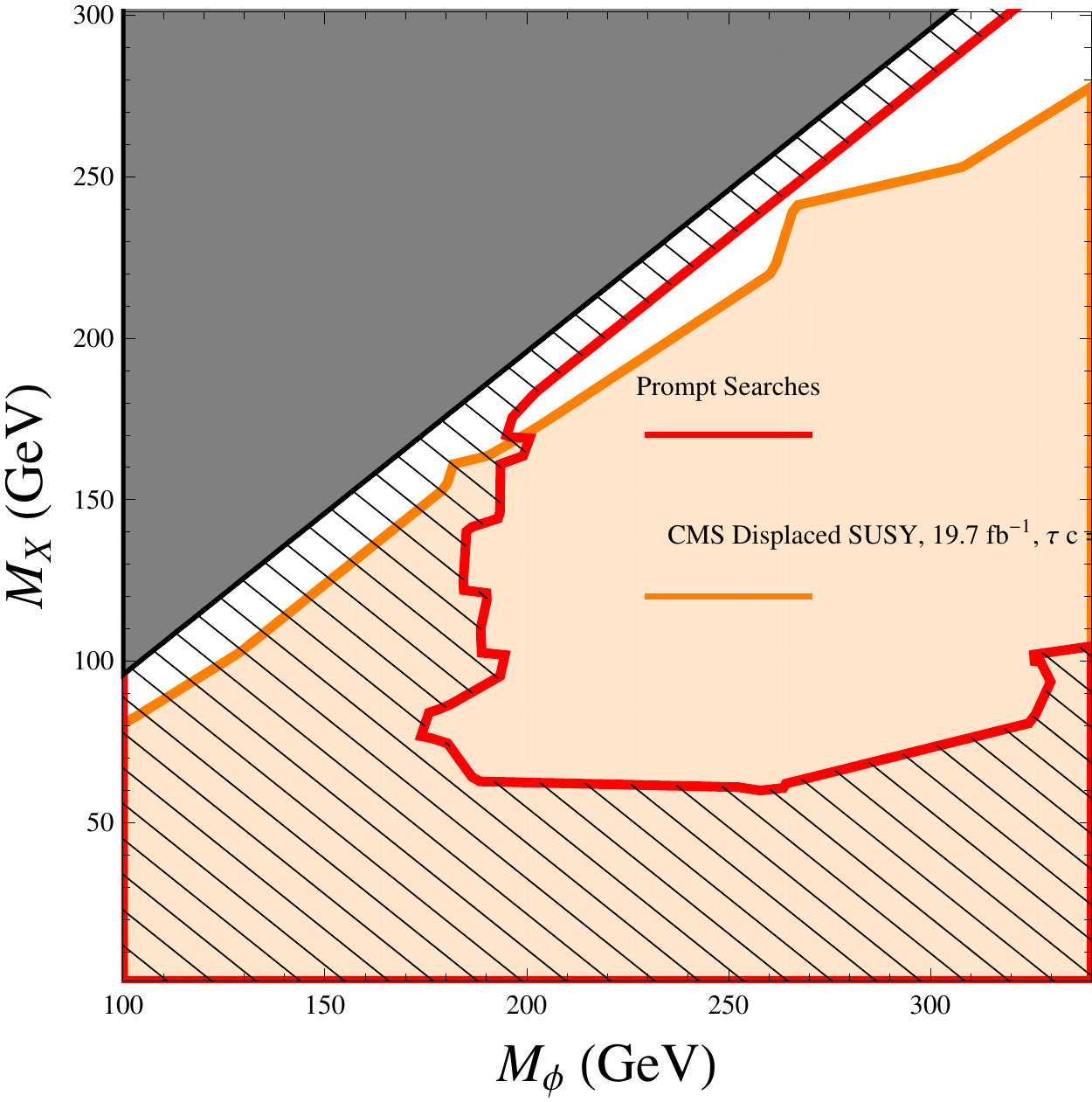}~
\caption{\small Allowed region (white) of parameter space consistent with the search by CMS for displaced leptons within a SUSY $R$-parity breaking scenario~\cite{Khachatryan:2014mea}. In Figure (a) (top panel), the colored scalar decays with a proper decay length of 0.1 mm and in (b) and (c) (bottom panel), with a decay length of 1 mm and 10 mm respectively. In all cases the hashed region bounded by the red solid line is excluded by prompt SUSY searches while the region bounded in orange is exclude by the CMS search for displaced leptons.}\label{fig:excl8}
 \end{figure*}

In Figure~\ref{fig:excl8}(a) (top panel) we show our exclusion region in the $m_{\phi}-m_{X}$ plane using the CMS results given in Table~\ref{tab:eventsEx} for a colored scalar with a proper lifetime of $0.1$ mm/$c$, while Figures~\ref{fig:excl8}(b) and (c) correspond to lifetimes of $1$ and $10$ mm/$c$ respectively. We observe stronger exclusions in regions where the SM background is low; that is, impact parameters between $0.1-2$ cm and decay lengths above $1$ mm. In particular, for lifetimes above $\gtrsim10$ mm, the CMS search for displaced leptons will reach beyond the prompt SUSY exclusions. Run II at the LHC with center of mass energies of $13$ TeV will extend the reach in areas that require a larger cross section or luminosity. 
Regions where a larger amount of missing transverse energy is present will require an analysis that incorporates signal regions using a hard cut on $\slashed{E}_{T}$ to reduce the SM background. This is particularly true for colored scalar masses between $\sim170-250$ GeV and a lifetime below $1$ mm$/c$ as well as dark fermion masses above $\sim 45-180$ GeV.

The authors in~\cite{Liu:2015bma} have carried out an in depth study to recast existing long-lived searches by the CMS and ATLAS collaborations for various general scenarios of SUSY with hadronic decays. In particular, they have explored the region with light scalar top quarks, motivated by naturalness, decaying to a top quark and a gravitino in the context of gauge mediation of SUSY breaking. In this scenario, a light stop decays to a gravitino a $W$ gauge boson and a $b$-quark through an off-shell top. They find that for a nearly massless gravitino, the displaced dijet analysis by CMS~\cite{CMS:2014wda} and the displaced muon plus tracks analysis by ATLAS~\cite{TheATLAScollaboration:2013yia} can exclude colored scalars with masses between $100$ and $200$ GeV with lifetimes all the way down to $0.1$ mm. However, for larger dark fermion masses, the gravitino in their context, the $4$-body decay mode of the colored scalar mode dominates leading to softer jets and a lower value of $H_{T}$. In addition, the analysis in~\cite{Liu:2015bma} places a very conservative bound on the lifetime of the colored scalar of $1$ mm when the decay is prompt, but as it was shown in the previous section, a lifetime below $1$ mm is not completely ruled out, in particular if the colored scalar has a mass above $\sim 200$ GeV and the dark fermion lies above $\sim80$ GeV.

In what follows, we analyze the LHC reach to our simplified model using $\sqrt{s}=13$ TeV. We propose a search strategy more sensitive to colored scalars with lifetimes below $1$ mm$/c$.

 \section{Long-lived Colored Scalars at LHC 13 TeV}\label{sec:LHC13}

In this section we discuss a very simple search strategy that can be used to probe long-lived colored scalars with a lifetime below $1$ mm$/c$ at the LHC. As we have shown in the previous section, for lifetimes above $1$ mm$/c$, the CMS search~\cite{Khachatryan:2014mea} rules out most of the parameter space up to colored scalar masses of $350$ GeV. This is due to the very little SM background present there. Even though, a lifetime below $\sim1.$ mm$/c$ is usually considered prompt, prompt searches have not yet ruled ruled out the entire parameter space and long-lived searches are still weak due to the large SM background. We thus outline how a search for leptons and missing energy can be used to compliment the exclusion carried out by prompt searches if we further apply a cut on the lepton's impact parameter. In addition, the strategy is simple as it can be used to completely eliminate the $t\bar{t}$ and QCD multijet backgrounds. We do this through a combination of both an upper bound on the sum of the jet $p_{T}$ in the event and a lower bound on the missing energy. The latter is used to suppress the QCD multijet background, as most of the MET arises from the misreconstruction of jets. In addition, we only trigger on a displaced muon arising form the decay of a long-lived particle within the inner detector. Other sources of backgrounds, such as diboson and $W+$jets are easily suppressed after requiring a transverse impact parameter for the muon above $0.01$ cm. 

\begin{figure*}[ht]
\centering
\includegraphics[width=7.0cm]{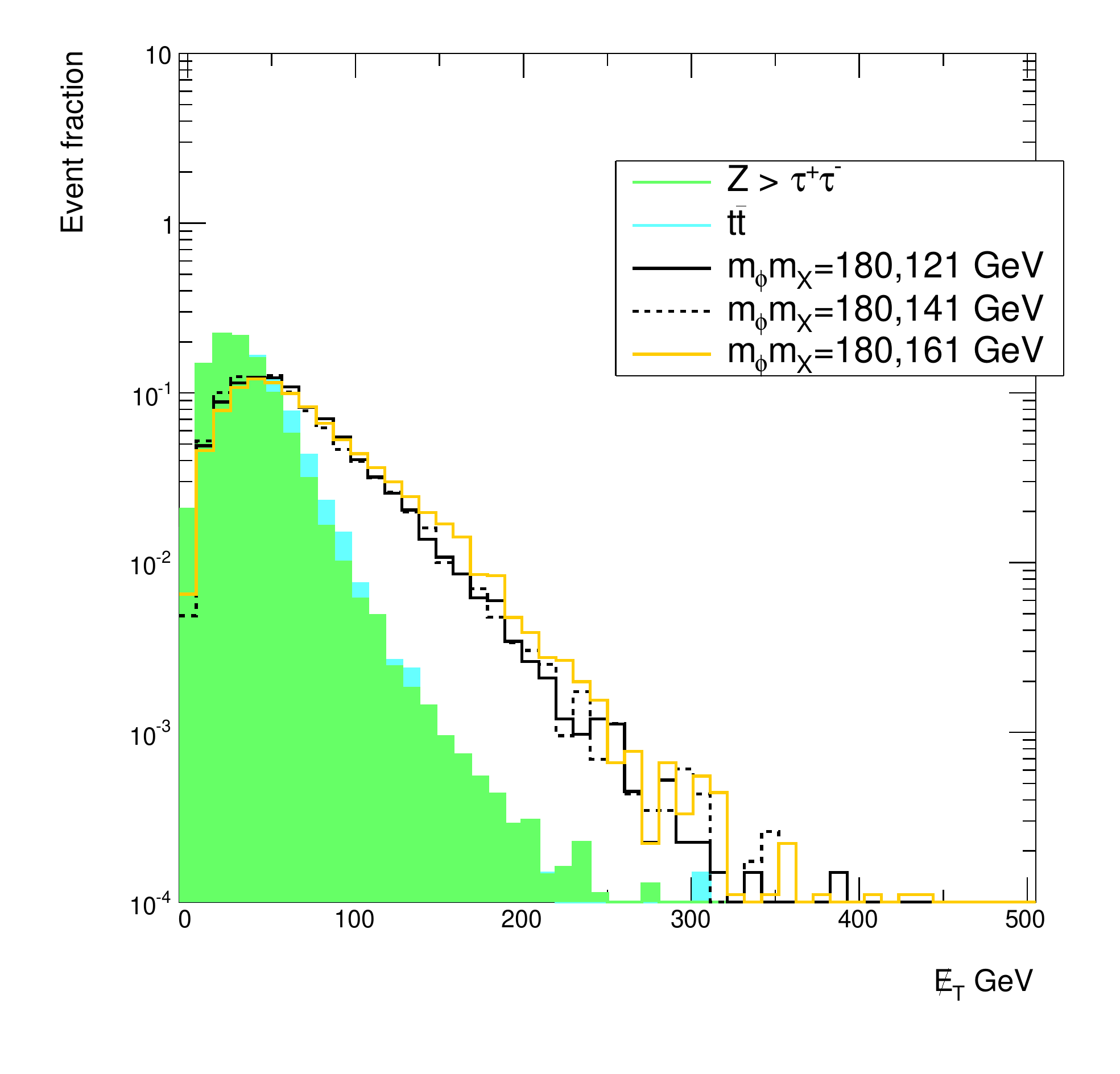}~
\includegraphics[width=7.0cm]{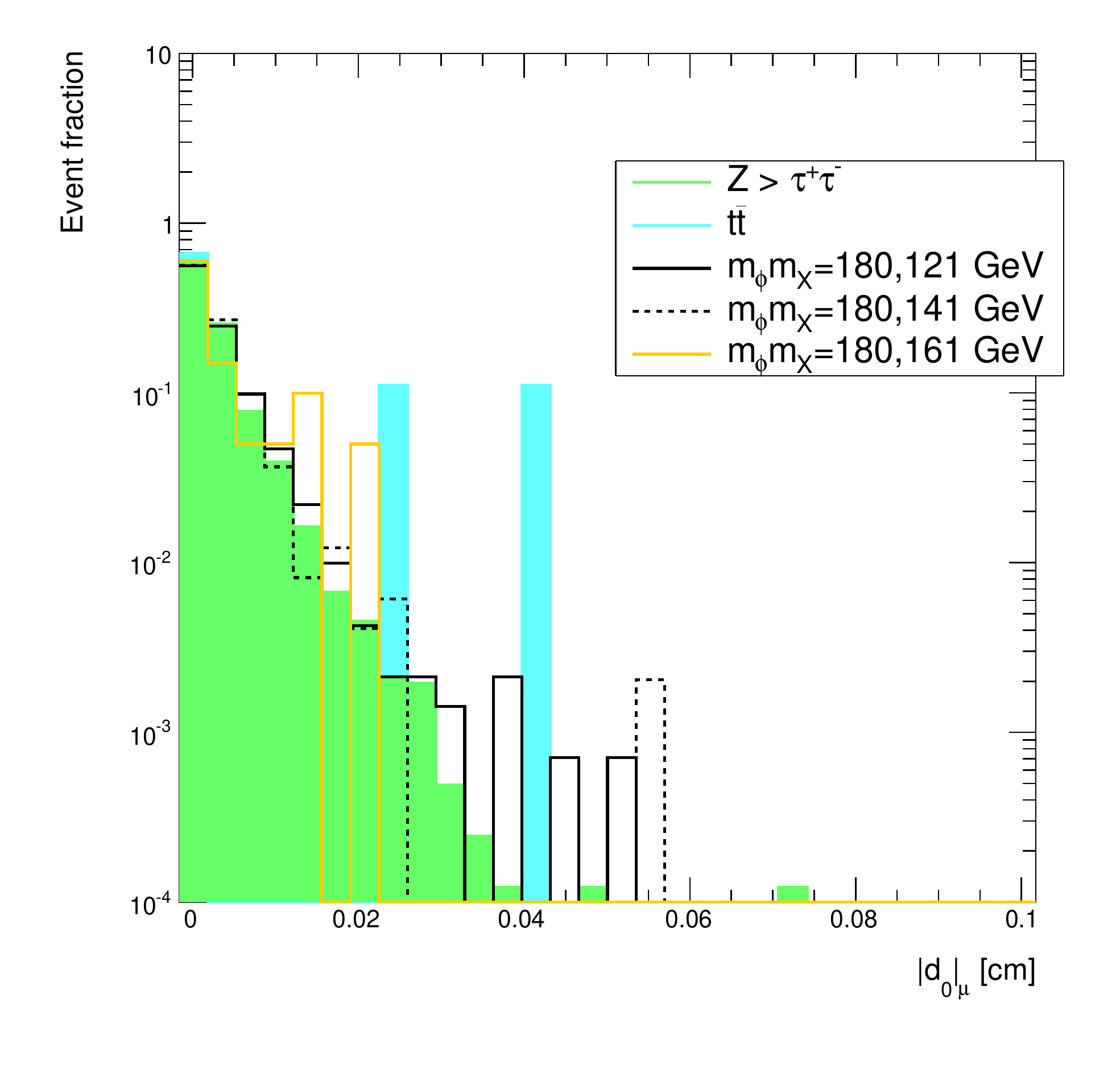}~
\caption{\small On the left, we show the fraction of events as a function of the missing transverse energy, $\slashed{E}_{T}$ while on the right we show the muon's transverse impact parameter. The figure corresponds to a colored electroweak-singlet scalar with decay length of $0.1$ mm.}\label{fig:fig7}
 \end{figure*}
We use MadGraph 5~\cite{Alwall:2011uj} in order to simulate the backgrounds and the pair production of a colored scalar with model files generated with FeynRules~\cite{Feynr}. Furthermore, as in the analysis described in Section~\ref{subsec:LLS8}, we only simulate the $Zj(Z\to\tau^{+}\tau^{-})$ and $t\bar{t}$ backgrounds. The parton showering and hadronization are carried out by Pythia~\cite{Sjostrand:2006za}, as well as the decays of the colored scalar which are implemented by a decay table that takes into account the 2-,3-, and 4-body decay branching fractions. In our analysis we use the Delphes~\cite{delphes} fast detector simulator to reconstruct jets with $p_{T}>10$ GeV using the anti-$k_{T}$ algorithm with parameter $\Delta R=0.5$, as well as to isolate and reconstruct tracks. The muon identification is performed by matching a track to a truth-level muon with $p_{T}>25$ GeV and $|\eta|<2.5$. We implement a muon reconstruction efficiency of $95\%$ and introduce an isolation criteria, $I_{l}$, by summing the $p_{T}$ of all tracks within a cone of size $\Delta R=0.4$ around the muon and requiring that 
\begin{equation}
 I_{\mu}=\frac{\sum_{i\ne l}p_{T,i}}{p_{T,l}} < 0.12.
 \end{equation}
Lastly, we reject events with $H_{T}>100$ GeV in order to suppress the QCD and $t\bar{t}$ backgrounds~\cite{TifSeq}, but advise that one may define different signal regions in $H_{T}$ to enhance the reach, especially at collider energies above $8$ TeV. The analysis is restricted to a muon's transverse impact parameters, defined in Equation~(\ref{eq:IP}), between $0.01-0.1$ cm to further reduce backgrounds where the lepton originates from the decay of a heavy flavor quark.
\begin{table}[ht]
\renewcommand{\arraystretch}{1.5}{
 \tabcolsep 2.2 pt
\small
\begin{center}
\begin{tabular}{|c|c|c|c|c||c| }
\hline
 $b$-jet & $\slashed{E}_{T}$ (GeV) &  $|d_{0}|_{\mu}$ (cm)  & $\sigma_{Zj}$ (pb) & $\sigma_{signal}$ (pb) & $\frac{N_{signal}}{\sqrt{N_{signal}+N_{background}}}$ \\
\hline
\hline
    - & $>50$  & $0.01<|d_{0}|<0.1$ & $0.074$ & $0.0023$ & $0.73,1.26,4.00$  \\
\hline
 - & $>100$  & $0.01<|d_{0}|<0.1$ & $7.88\times10^{-4}$ & $5.28\times10^{-4}$ & $0.70,1.21,3.83$  \\
 \hline
  1 & $>50$  & $0.01<|d_{0}|<0.1$ & $7.88\times10^{-4}$ &  $9.80\times10^{-4}$  & $2.00,3.47,10.97$   \\
  \hline
\end{tabular}
\end{center}
}
\caption{\small Signal and background cross sections after a series of cuts for a colored scalar with mass, $m_{\phi}=200$ GeV and a dark fermion of mass, $m_{X}=150$ GeV. The average lifetime for the scalar is of $0.1$ mm/$c$. In the last column we show the significance for luminosities of $10,30$ and $100$ fb$^{-1}$ with an overall upper bound on the sum of the jet $p_{T}$ given by $\sum p_{T,j}<100$ GeV. } \label{tab:Table2}
\end{table} 
In Figures~\ref{fig:fig7}(a) and~\ref{fig:fig7}(b) we show the $\slashed{E}_{T}$ and $|d_{0}|_{\mu}$ distributions for the $Zj(Z\to\tau^{+}\tau^{-})$ and $t\bar{t}$ backgrounds together with the signal from a colored scalar with a proper lifetime of $0.1$ mm/$c$. In the figure, the black solid line corresponds to $m_{\phi}=180$ GeV and $m_{X}=121$ GeV. The dashed back line, solid yellow line correspond to values of $m_{\phi} (m_{X})=180~(141),180~(161)$ GeV respectively.

\begin{figure*}[ht]
\centering
\includegraphics[width=7.0cm]{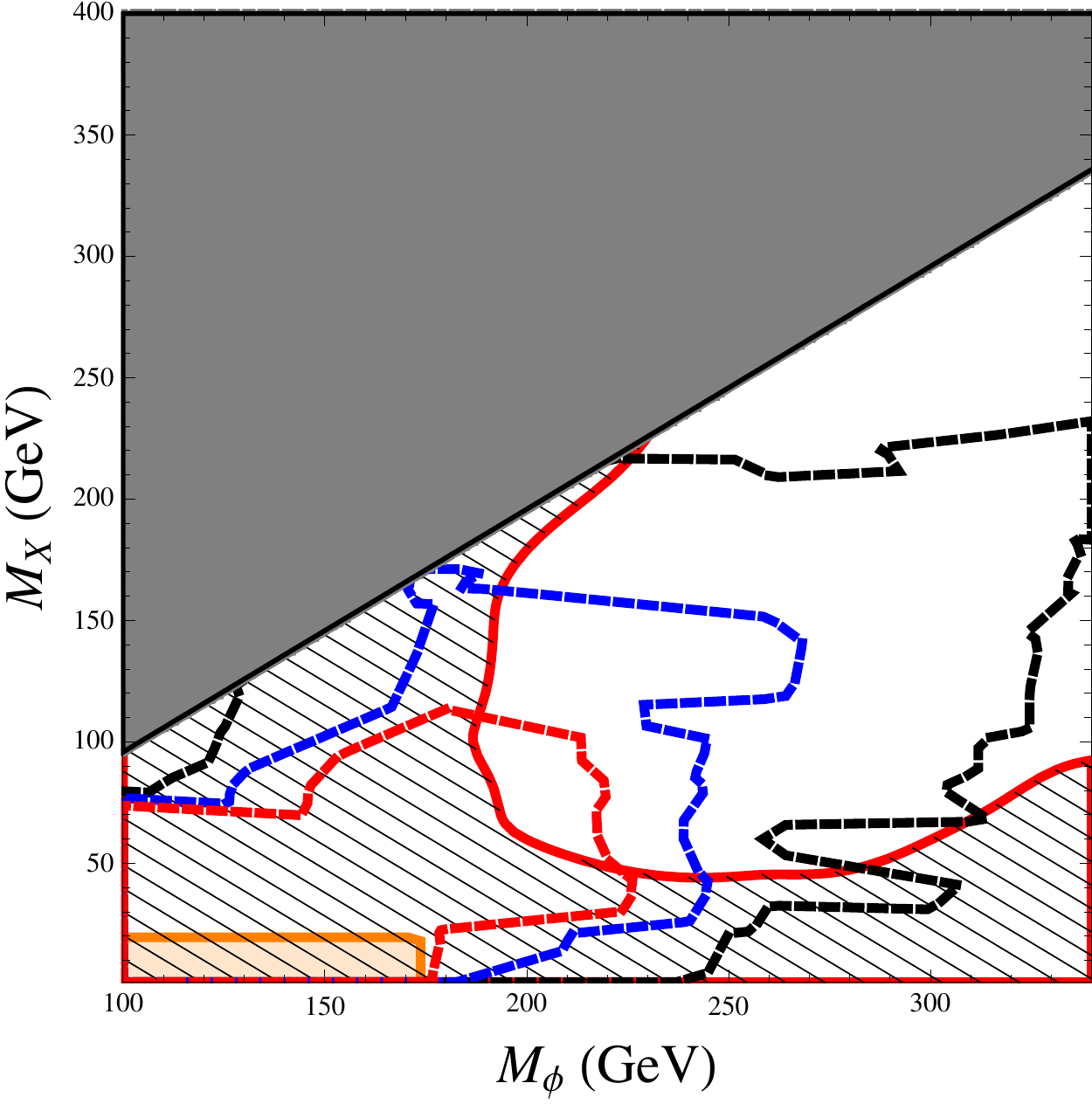}~
\caption{\small LHC reach at $13$ TeV for a signal consisting of a displaced muon together with a $b$-jet and missing transverse energy associated with the dark fermion, $X$ and corresponding to a colored scalar lifetime of $0.1$ mm/$c$. The regions bounded by the dashed black, blue and red lines depict values of $s>5$ for luminosities of $300,30$ and $10$ fb$^{-1}$ respectively. The hashed region bounded by the red solid line is excluded by prompt SUSY searches while the region in light orange is excluded by the CMS long-lived search discussed in Section~\ref{subsec:LLS8} .} \label{fig:fig8}
 \end{figure*}
In Table~\ref{tab:Table2} we show the LHC reach at $13$ TeV for a colored scalar with mass $m_{\phi}=200$ GeV and a lifetime of $0.1$ mm$/c$ together with a dark fermion with mass, $m_{X}=150$ GeV. To probe the sensitivity to this simplified model we define the significance of a signal using the following statistical estimator 
\begin{equation}
s=\frac{N_{s}}{\sqrt{N_{s}+N_{b}}},
\end{equation}
where $N_{s}$ and $N_{b}$ denote the number of signal an background events. Most of the variation depends mostly on the amount of missing momentum and the ability to $b$-tag the event (to further suppress the $Z\to\tau\tau$ in exchange of a significant drop in the $b$-tagging efficiency). From Table~\ref{tab:Table2} we observe that in the degenerate window; the most important cut is on the number of $b$-jets; since this region of parameter space is already characterized by a large amount of $\slashed{E}_{T}$. In the non-degenerate window, a stronger reach can be achieved by keeping events with lower $\slashed{E}_{T}$. Therefore; we would like to point out that $b$-tagging is only relevant in regions where small MET is required to increase the number of signal events and to significantly suppress the background. Requiring two $b$-jets will only destroy any possibility to extract the signal. In Figure~\ref{fig:fig8} we show the entire light colored scalar region in the $m_{\phi}-m_{X}$ plane that the LHC Run II may be able to probe with $10$, $30$ and $300$ fb$^{-1}$ respectively.

\section{Realizations}\label{sec:realizations}



The framework studied in this work fits naturally into the light stop window scenario studied in~\cite{Delgado}. In that work, the MSSM is consistent with collider data for stop masses above $200$ GeV, naturally emerges from renormalization group (RG) evolution, predicts the correct dark matter relic abundance and it is consistent with flavor constraints. This scenario can be accommodated through a light mostly right-handed stop with mass between $200-400$ GeV (which corresponds to $\phi$ in this work), a heavy mostly left-handed stop, a gluino with mass below $1.5$ TeV and a light neutralino (here, $X$) with nearly degenerate with the light stop.

Given an LSP neutralino with a mass very close to the stop mass, $\Delta m \,\equiv \, m_{\tilde{t}_1} - m_X \approx 20-50\, {\rm GeV}$, besides the two-body flavor-violating decay $\tilde{t}_1\to c \,X$, the four body decays are also relevant. For small $\Delta m$,
\begin{eqnarray}
&&\Gamma (\tilde{t}_1\to c\,X) =  100 {\rm \, cm}^{-1}\,\left(\frac{\theta_{tc}}{10^{-5}} \right)^2\left( \frac{\Delta m}{30\,{\rm GeV}} \right)^2 \left(\frac{400\,{\rm GeV}}{m_{\tilde{t}_1}} \right), \\
&&\Gamma (\tilde{t}_1\to X\,b\,\ell^+ \nu_\ell )  =  28  {\rm \, cm}^{-1}\,\left( \frac{\Delta m}{30\,{\rm GeV}} \right)^8 \left(\frac{400\,{\rm GeV}}{m_{\tilde{t}_1}} \right),\\
&& \Gamma (\tilde{t}_1\to X\,b\,b\,\bar{d})  \approx \Gamma (\tilde{t}_1\to X\,b\,c \bar{s} )  \approx 3 \Gamma (\tilde{t}_1\to X\,b\,\ell^+ \nu_\ell ),
\end{eqnarray} where $\theta_{tc}$ is the stop-scharm mixing angle, which in \cite{Delgado} is estimated to be of order $\mathcal{O}(10^{-5})$. We see that these decays yield decay lengths between $\ell_{\tilde{t}} \approx 0.01\,-\,0.1\,{\rm mm}$. However, the stop can be longer lived if one relies on suppressing $\theta_{tc}$ below $10^{-5}$, yielding displaced vertices between $\sim0.1-0.5$ mm for a $200$ GeV stop mass.
 \begin{figure*}[ht]
\centering
\includegraphics[width=7.0cm]{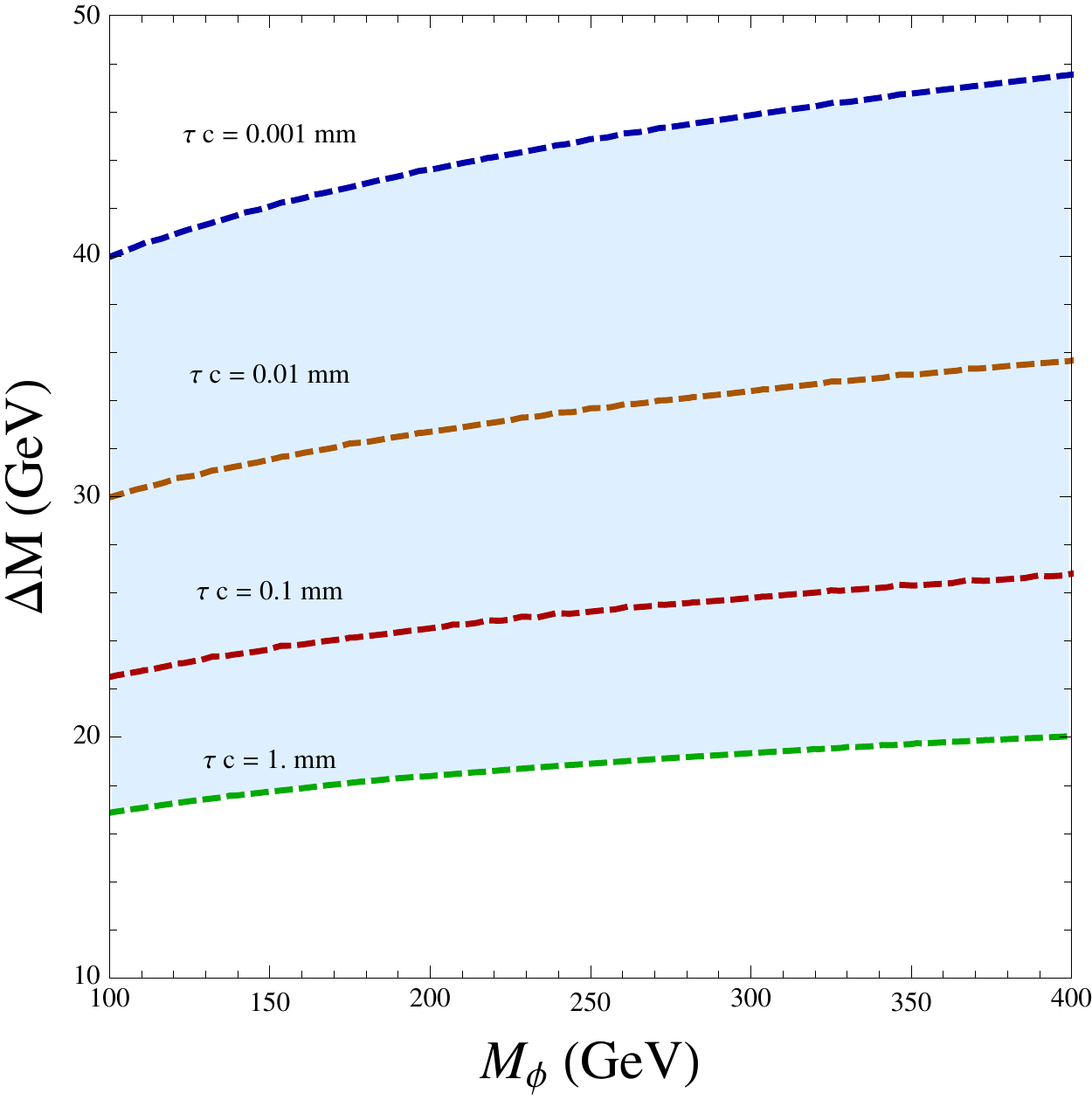}~
\includegraphics[width=7.0cm]{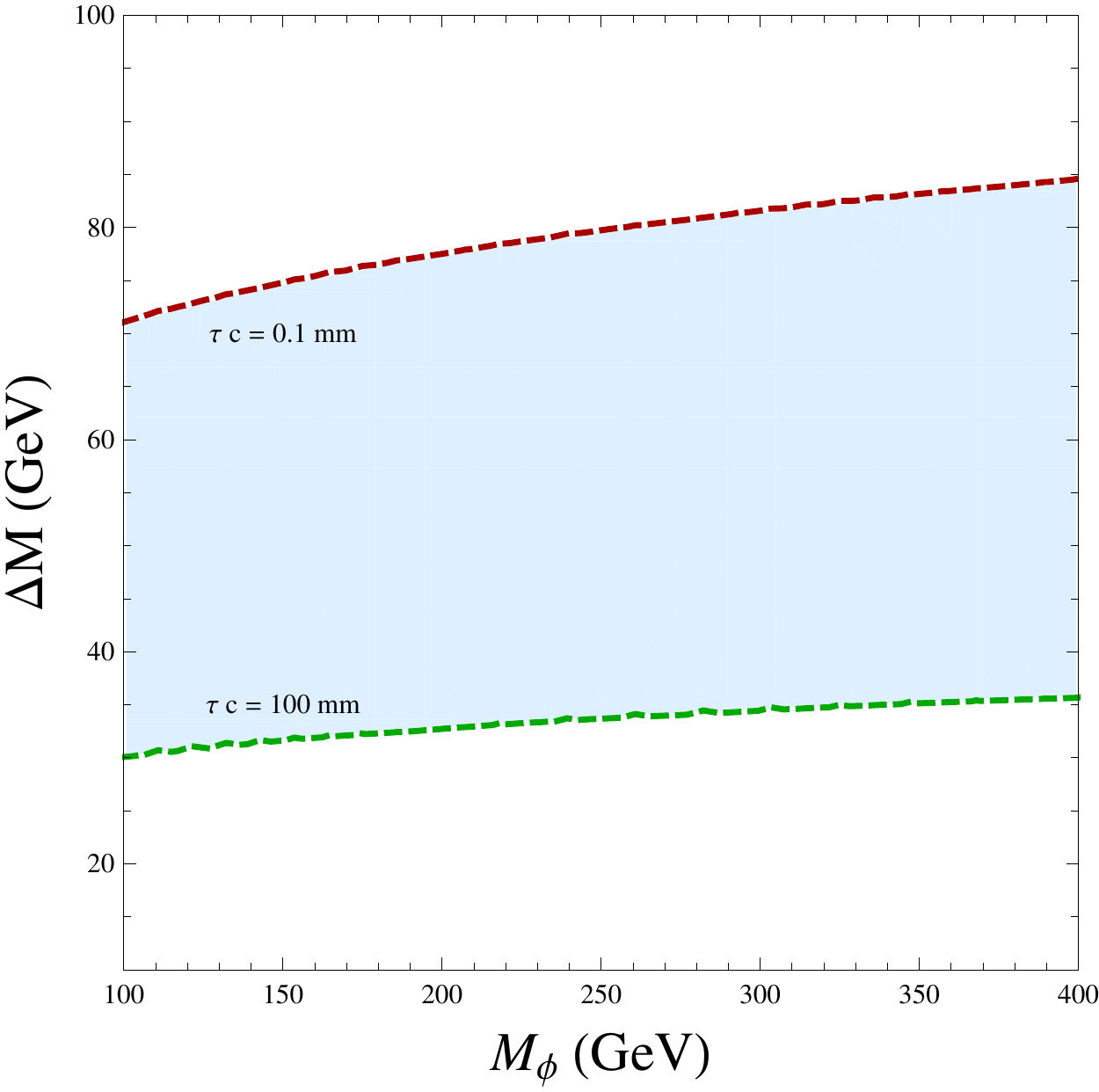}~
\caption{\small Region of parameter space in the light-stop window of the MSSM (left) corresponding to scalar top decay lengths between $0.001$ and $1$ mm. On the right we show a scenario where the width of the scalar top quark is further suppressed by an ${\cal O}\left(0.01\right)$. Within this scenario, decay lengths between $1$ and $10$ mm are accessible. }\label{fig:Fig14}
 \end{figure*}
Furthermore, in Figure~\ref{fig:Fig14}(a), we show the shaded region consistent with decay lengths in the range $0.001-1$ mm in the absence of a flavor violating decay mode. Larger lifetimes will require an additional suppression of the stop $4$-body decay mode. This suppression may exist in extensions of the MSSM where the lightest neutralino is mostly a SM gauge singlet. In Figure~\ref{fig:Fig14}(b), the shaded regions denote decay lengths in the range $0.1-100$ mm after suppressing the stop-neutralino coupling by an ${\cal O}\left(0.01\right)$ coupling. This region of parameter space is then a driving force motivating an analysis for displaced vertices such as the one discussed in Section~\ref{sec:LHC13} to further probe this light stop window of the MSSM for dark fermion masses above $\sim50$ GeV.

\section{Conclusions}\label{sec:discussion}
We have studied a simplified scenario where a colored electroweak-singlet scalar couples to a dark sector through the right-handed top quark. Within our framework the colored scalar can decay after traveling through the inner detector leading to displaced leptons and missing energy. The amount of missing energy strongly depends on the spectrum of the model, but it can be sufficiently large to escape current bounds on displaced vertices that mostly target the parameter space of $R$-parity violating scenarios of the MSSM.

After implementing a recent search by the CMS collaboration we observe that a color scalar with lifetime in the range $\sim0.1-1$ mm$/c$ and masses between $\sim200-350$ GeV may have escaped detection. We have proposed a search strategy at the $13$ TeV LHC that can probe a large enough amount of the parameter space with as little as $10$ fb$^{-1}$ by implementing a cut on the lepton's impact parameter and requiring a large amount of missing transverse energy. However, a statistical significant probe will require between $100-300$ fb$^{-1}$.

In addition, we observe that our simplified model naturally fits into light-stop window scenarios of supersymmetric extensions of the SM, where we observe that flavor violating mixing angles below $10^{-5}$ can lead to purely displaced vertices between $0.1-10$ mm for colored scalar masses between $100-250$ GeV. This is incredibly exciting since it opens the possibility that a slightly long-lived right-handed stop could be spotted at the LHC.

\section*{Acknowledgements}

The authors thank Walter Tangarife for his critical feedback throughout the progress of this work. A.D.P would like to thank Travis A.W. Martin and David Morrissey for useful discussions and essential feedback regarding the progress of this work. A.D.P is supported in parts by the National Science and Engineering Council of Canada. A.S has been partially supported by ANPCyT under grant No. PICT-PRH 2009-0054 and by CONICET.

\end{document}